\documentclass[10pt, conference, compsocconf]{IEEEtran}

\usepackage{float}
\floatstyle{ruled}
\usepackage{graphicx}
\usepackage{amsmath}
\usepackage{verbatim}
\newfloat{program}{thp}{lop}
\floatname{program}{Listing}
\usepackage{multirow}
\usepackage{subfigure}
\usepackage{eso-pic}

\AddToShipoutPicture*{\small \sffamily\raisebox{1.8cm}{\hspace{1.8cm}978-1-4577-0351-5/11/\$26.00 \copyright2011 IEEE}}

\begin{document}

\title{Exploiting Temporal Complex Network Metrics in Mobile Malware Containment}

\author{\IEEEauthorblockN{John Tang}
\IEEEauthorblockA{Computer Laboratory\\
University of Cambridge\\
jkt27@cam.ac.uk
}
\and
\IEEEauthorblockN{Cecilia Mascolo}
\IEEEauthorblockA{Computer Laboratory\\
University of Cambridge\\
cm542@cam.ac.uk
}
\and
\IEEEauthorblockN{Mirco Musolesi}
\IEEEauthorblockA{School of Computer Science\\
University of St. Andrews\\
mirco@cs.st-andrews.ac.uk}
\and
\IEEEauthorblockN{Vito Latora}
\IEEEauthorblockA{Dipartimento di Fisica\\
University of Catania\\
latora@ct.infn.it}
}

\maketitle

\begin{abstract}
Malicious mobile phone worms spread between devices via short-range Bluetooth contacts, 
similar to the propagation 
of human and other biological viruses.  
Recent work has employed models from epidemiology and complex networks to analyse the spread of malware and the effect of  patching specific nodes. These approaches have adopted a static view of the mobile networks, i.e., by aggregating all the edges that appear over time, which leads to an approximate representation of the real interactions: instead, these networks are inherently dynamic 
and the edge appearance and disappearance are highly influenced by the ordering of the human contacts, something which is not captured at all by existing complex network measures.

In this paper we first study how the blocking of malware propagation through immunisation of key nodes (even if carefully chosen through static or temporal betweenness centrality metrics) is ineffective: this is due to the richness of alternative paths in these networks. 
Then we introduce a \textit{time-aware} containment strategy that \textit{spreads} a patch message starting from nodes with high temporal closeness centrality and show its effectiveness using three real-world datasets.  Temporal closeness allows the identification of nodes able to reach most nodes quickly: we show that this scheme 
reduces the cellular network resource consumption and associated costs, achieving, at the same time, complete containment of malware in a limited amount of time.   
\end{abstract}
\vspace{-8pt}
\section{Introduction}
\label{sec:introduction}

Smartphones are not only ubiquitous, but also an essential part of life for many people who carry such devices through their daily routine.  It comes at no surprise then that recent studies have shown that the mobility of such devices mimic that of their owners' schedule~\cite{eagle_reality_2006,wang_understandingspreading_2009}. 
This fact constitutes an opportunity for devising efficient protocols and applications, but it also represents an increasing security risk: as with biological viruses that can spread from person to person, mobile phone viruses can also leverage the same social contact patterns to propagate via short-range wireless radio such as Bluetooth and WiFi.  
For example, when security researchers downloaded \textit{Cabir}~\cite{cabir} 
-- a proof-of-concept mobile worm -- 
for analysis, they 
discovered the full risk 
 as it broke loose, replicating from the test device to external mobile phones. 
This 
prompted the need for specially radio shielded rooms to securely test such malicious code~\cite{f-secure_2005}.

Until recently though, 
mobile malware has been developed only for proof-of-concept experiments with very limited and non malicious effects on users
~\cite{stringer_six_2008,schipka_dollars_2009}.
However, the immense popularity and improvements in smartphone technology have attracted the attention of a growing number of attackers. 
In particular, increasing economic incentives have been the motivation of more recent exploits, for example stealing private data such as phone contacts~\cite{liu_symbos_pbstealer}; 
transferring call credit to other accounts \cite{kaspersky_lab_kaspersky_2009};
and 
traditional 
exploits such as premium rate number dialling~\cite{terdial_2010}.

Unlike desktop computers mobile malware can spread through both short-range radio (i.e., Bluetooth and WiFi) and long-range communication (i.e., SMS, MMS and email)~\cite{leavitt_mobile_2005}.  Long-range malicious traffic can potentially be contained by the network operator by scanning every message against a database of known malware~\cite{van_ruitenbeek_quantifyingeffectiveness_2007}, 
however, short-range propagation might fall under the radar of centralised service providers: effective schemes to defend against short-range mobile malware spreading are necessary.
Also, while a global patching of the devices through cellular connectivity is the natural solution and is in theory possible, in practice,  
there are potential constraints with respect to the cellular network capacity and server bandwidth~\cite{enck_exploiting_2005}.

Being highly correlated with human contacts, understanding how such malware propagates requires an accurate analysis of the underlying time-varying network of contacts amongst individuals.
State-of-the-art solutions on mobile malware containment have ignored two important temporal properties: firstly, the time order, frequency and duration of contacts; and secondly, the time of day a malicious message starts to spread and the delay of a patch~\cite{zhu_social_2009,zyba_defending_2009}. Instead, we argue that the temporal dimension is of key importance in devising effective solutions to this problem.

With this in mind, the focus of this study is to investigate the effectiveness of two containment strategies based on targetting key nodes, taking into account these temporal characteristics. 
We firstly investigate a traditional strategy, inspired by studies on error and attack tolerance of networks~\cite{albert_error_2000}, exploiting a static and a time-aware enhanced version of \textit{betweenness centrality} which provide the best measure of nodes that mediate or bridge the most communication flows.
According to this strategy the nodes that act as mediators are patched to \textit{block} the path of a malicious message. However, due to temporal clustering and alternative temporal paths, in most cases, such strategies merely \textit{slow} the malware and does not \textit{stop} it. 
In other words, a scheme based solely on immunisation of key nodes is not sufficient, instead {\em quick spreading} of the patch is necessary for most networks.
We propose a solution based on local \textit{spreading} of patches through Bluetooth, i.e., exploiting the same mechanism used by the malware itself. The key issue in this approach is to select the right nodes as starting points of the patching process.
Temporal betweenness only provides a quantitative measure of the number of communication paths over time that go through a certain node and it proves to be sub-optimal metric for this. A metric capable of identifying nodes that can reach a large quantity of other nodes quickly is needed. Our choice fell on \textit{temporal closeness centrality} which ranks nodes by the speed at which they can disseminate a message to all other nodes in the network. We show that this strategy can reduce the cellular network resource consumption and associated costs, achieving at the same time a complete containment of the malware in a limited amount of time.

In the following sections, we will first introduce some preliminary definitions related to temporal graph analysis and metrics and then present a detailed study of our proposed containment scheme using real-world traces.

\section{Temporal Networks}
\label{sec:temporal_graphs}

Temporal graphs have recently been proposed \cite{tang_temporal_2009,grindrod_evolving_2009} to study real dynamic datasets, with the intuition that the behaviour of dynamic networks can be more accurately captured by a sequence of \textit{snapshots} of the network topology as it changes over time instead of using a representation whereby all the contacts are aggregated into a single static graph.
From this, temporal versions of shortest path \cite{tang_temporal_2009} have demonstrated that, since static analysis ignores time ordering of contacts, static shortest paths {\em overestimate} the available links and {\em underestimate} the actual shortest path length.

\begin{figure}[t]
  \begin{center}
    \subfigure[Temporal Graph]{\label{fig:tg-a}\includegraphics[scale=0.39, trim = 15mm 8mm 0mm 0mm, clip]{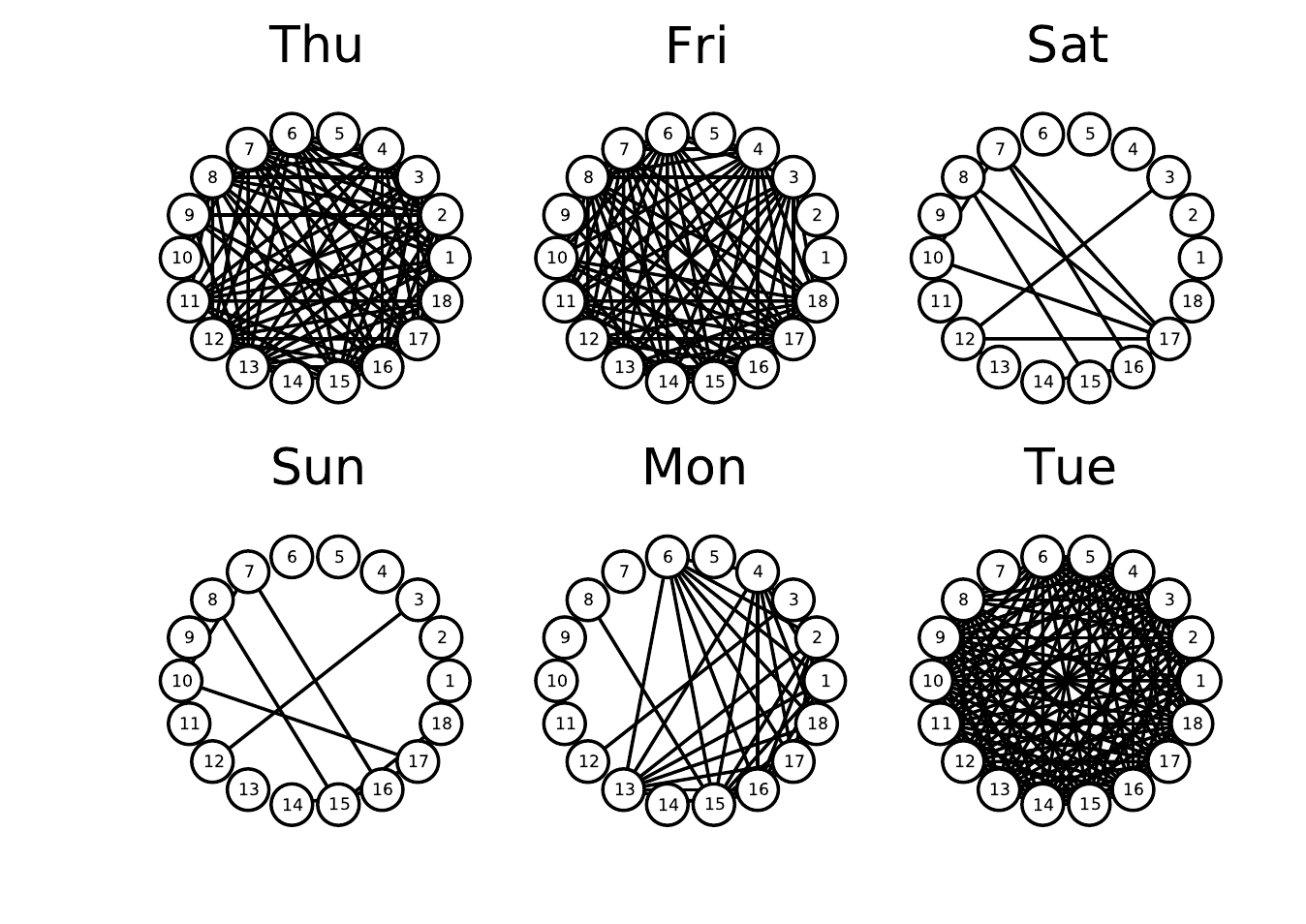}}
    \subfigure[Static Graph]{\label{fig:tg-b}\includegraphics[scale=0.39, trim = 10mm 8mm 15mm 15mm, clip]{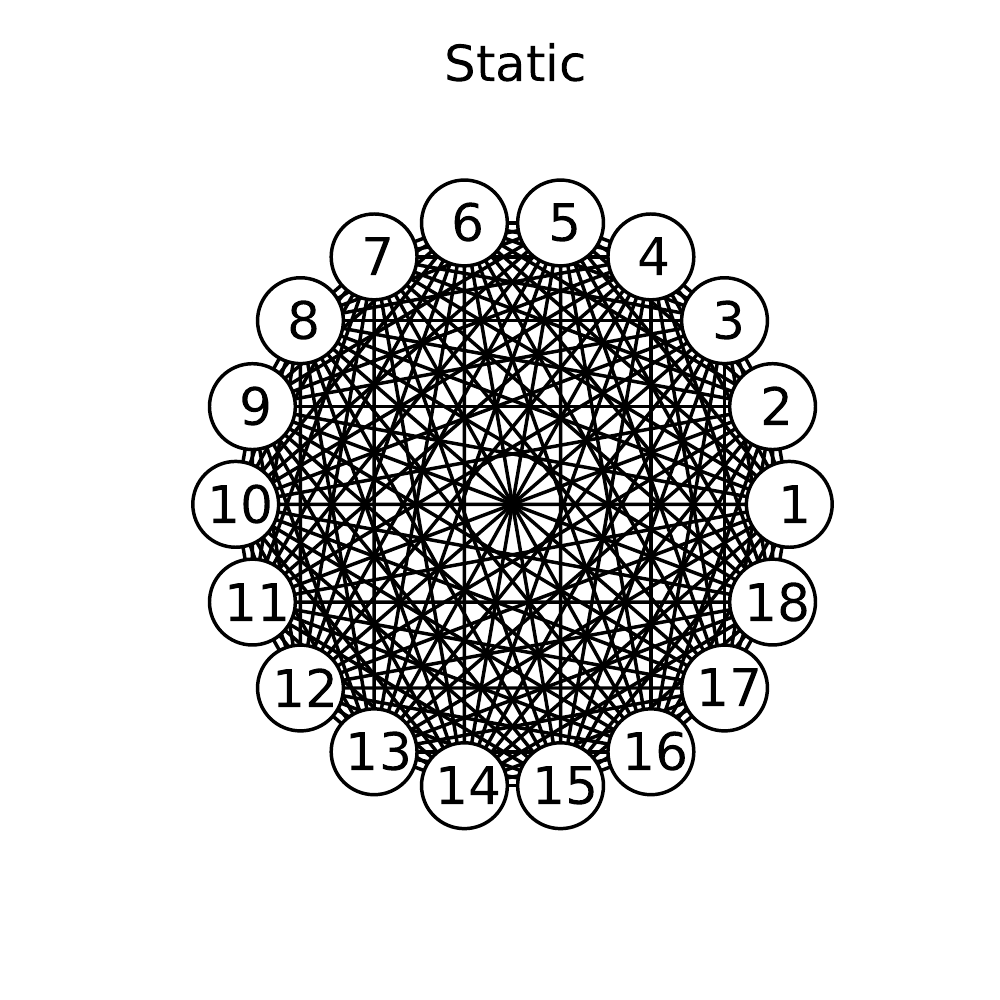}} \\
  \end{center}
 \vspace{-11pt}
\caption{(a) Temporal Graph showing contacts using 24-hour windows and (b) aggregated static graph for the {CAMBRIDGE} dataset.  Nodes represent devices; two nodes are linked if there was a Bluetooth contact within that 24-hour window.}
  \label{fig:tg}
\vspace{-14pt}
\end{figure}

We now provide a brief overview of these concepts in relation to the problem of designing effective malware containment schemes. Since Bluetooth radio can only handle \textit{uni-directional} transmissions (from the scanning device to the scanned device) we define a \textit{directed} temporal graph 
which can be thought of as an ordered sequence of directed graphs\footnote{This does not lose generality of bi-directional communication since transmissions can still be reciprocated during the same encounter.}.  
A state of the network topology is calculated by aggregating all the directed edges that appear inside a certain time window. An example, using a dataset of contacts among students and staff at Cambridge, is given in Figure~\ref{fig:tg} (more details about the dataset are provided in Section~\ref{sec:datasets})\footnote{Direction of edges have been removed for clarity.}:
Figure \ref{fig:tg-a} shows a temporal graph with a sequence of six graphs,
each of them representing the contacts among devices in a time window of
24-hours. The corresponding aggregated static graph (which
reports all the links amongst nodes, without any information about
time) is shown in 
Figure \ref{fig:tg-b}. 
The static graph misses the circadian rhythms that can instead be observed in the temporal graph.  
Also note that the high density of links within the static graph, which we will see later, contributes to problems in discriminating between important nodes for the calculation of static centrality; instead a temporal graph is required to capture the rich temporal information of the interaction patterns.

To give an intuition as to why temporal graphs and temporal paths are necessary, consider the temporal graph and associated aggregated static graph in Figure \ref{fig:tg2}. 
If we consider a shortest path from node $A$ to $F$, according to the static graph there is a 2-hop path $(A,C,F)$, when in fact taking into account the \textit{time ordering} of contacts in the temporal graph, we see that such a path does not exist in reality; instead, the actual shortest path is of 3-hops $(A,C,E,F)$.

More formally, given a real-world contact trace starting at $t_{min}$ and ending at
$t_{max}$, the directed temporal graph ${\cal G}^w(t_{min},t_{max})$ is defined as the ordered sequence of graphs
($G_{0}$, $G_{2}, \ldots, G_{T-1}$) where
$T=((t_{max}-t_{min})/w) = |{\cal G}^w(t_{min},t_{max})|$ is the number of graphs in the sequence and $w$ is
the size of each time window expressed in some time units (e.g.,
seconds or hours).
There exists a directed link from $i$ to $j$ in
$G_{T}$ if there is a contact from $i$ to $j$ during the time interval $[(t_{min} + (w\times T)) , (t_{min} + (w\times (T+1))) )$. 
All graphs in the temporal graph have the same set of nodes $V$.

From this a \textit{temporal path} starting at $i$ and finishing at $j$ can be defined over ${\cal G}^w(t_{min},t_{max})$ as a sequence of $k$ hops via a distinct node
$n_k^{W_k}$ at time window $W_k$:
\begin{equation}  \label{eq:p}
p_{ij}^{h} = (n_1^{W_0}, \ldots, n_{k}^{W_k})
\end{equation}  
where $i=n_1$, $j=n_k$, 
node $n_k$ is passed a message at time window
$W_k \geq W_{k-1}$, $0 \leq W_k < T$
and $h$ is the maximum hops through which a message is replicated
within the same window.  
Subsequent definitions implicitly set $h = 1$, since higher values of $h$ lead to similar performance of the containment schemes.
We call $Q_{ij}$ the set of all temporal paths between nodes $i$ and $j$.
If a temporal path from $i$ to $j$ does not exist i.e. $Q_{ij}=\emptyset$, we say
that $(i,j)$ is a \textit{temporally disconnected node pair}, and we
set the distance $d_{ij}= \infty $. 

Using the function $D(p_{ij})=(w\times W_k)$ which returns
the real delivery time (at window $W_k$) for the given path relative to $t_{min}$, the \textit{shortest temporal path length} 
is defined as:
\begin{equation}  \label{eq:d}
  d_{ij} = min(D(q_{ij})), \forall q_{ij} \in Q_{ij}
\end{equation}  
From this we define the set $S_{ij}$ of \textit{shortest temporal paths} between
$i,j$ as:
\begin{equation}  \label{eq:S}
   S_{ij} = \{p_{ij} \in Q_{ij} \mid D(p_{ij})=d_{ij} \} 
\end{equation}  

We define the \textit{temporal efficiency} $E_{ij}$ between nodes $i$ and $j$ for the time interval $t_{min}$ to $t_{max}$ as:
\begin{eqnarray}  \label{eq:E}
E_{{ij}}(t_{min},t_{max}) &=& \frac{1}{d_{ij}(t_{min},t_{max})+1}
\end{eqnarray}  
We can then define 
the average \textit{temporal efficiency} $E$ 
as:
\begin{eqnarray}
& \displaystyle 
E(t_{min},t_{max}) = \frac{1}{N(N-1)} \sum_{\substack{i,j \in V \\ i\neq j}} E_{{ij}}(t_{min},t_{max})  \label{eq:Eglob} ~~
\end{eqnarray}
For brevity we shall also refer to this as \textit{efficiency}. Efficiency naturally handles disconnected node pairs since it gives us the harmonic mean of delays between all node pairs.

\section{Temporal Centrality Metrics for Malware Containment}
\label{sec:temporalcentrality}
\begin{figure}[t!]
	\center
	\includegraphics[scale=0.4, trim = 0mm -5mm 0mm 0mm, clip]{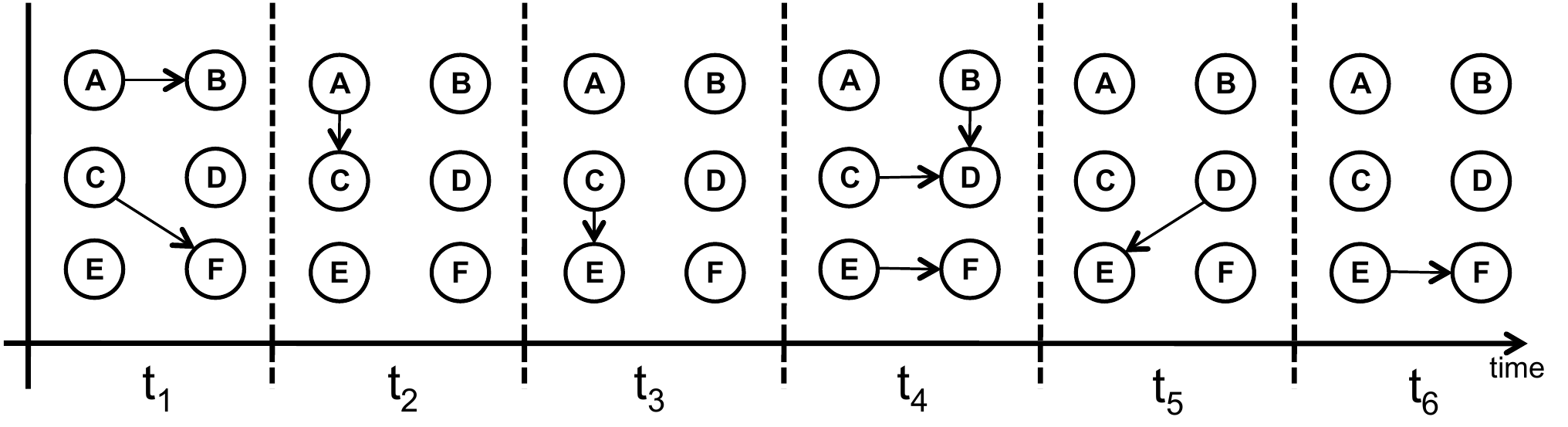}
	\includegraphics[scale=0.4, trim = 0mm 0mm 0mm 0mm, clip]{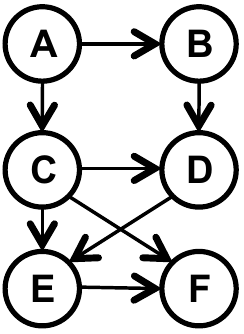}
\vspace{-8pt}	
\caption{Example directed Temporal and aggregated static graph.}\label{fig:tg2}
\vspace{-10pt}
\end{figure}
Let us consider a simple scenario where a person receives a malicious message on their device in the early hours of the morning and the malicious program replicates itself to any devices it meets during the day, for example at work and while socialising in the evening. 

\begin{figure}[t!]
  \begin{center}
    \subfigure[]{\label{fig:tgpaths-a}\includegraphics[scale=0.4, trim = 0mm 0mm -10mm 7mm, clip]{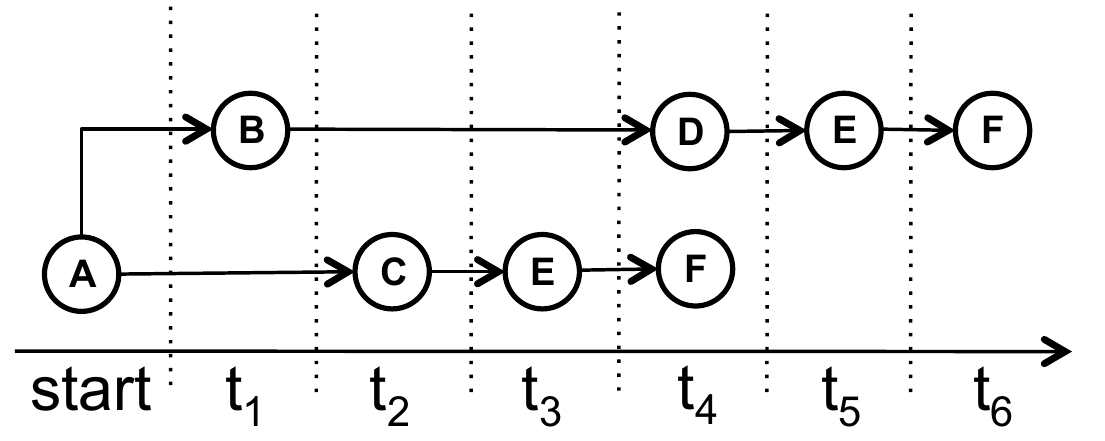}}
    \subfigure[]{\label{fig:tgpaths-b}\includegraphics[scale=0.4, trim = 0mm 0mm 3mm 7mm, clip]{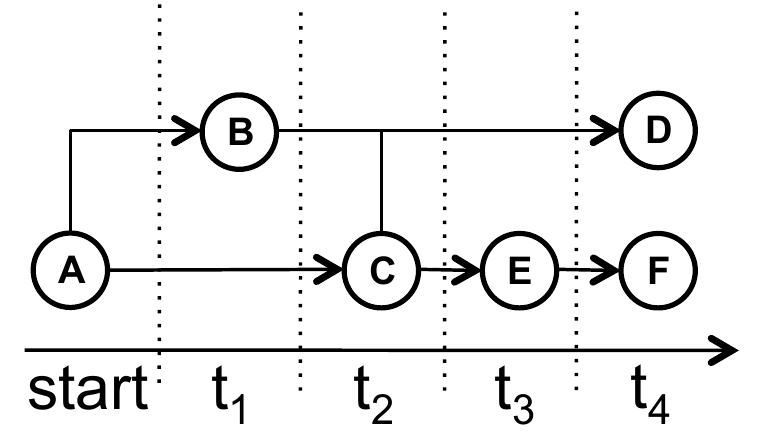}} \\
  \end{center}
 \vspace{-13pt}
  \caption{(a) Two temporal paths from $A$ to $F$.  (b) Temporal minimum spanning tree with source $A$ and shortest temporal paths to all other nodes.}\label{fig:tgpaths}
  \label{fig:path3}
 \vspace{-15pt}
\end{figure}

A simple strategy 
consists of immunising only the nodes 
which mediate the most communication flows.
{\em Betweenness centrality} metrics have been devised for static complex network graphs to measure this quantity~\cite{wasserman_social_1994} and we have extended this measure to incorporate the temporal dimension~\cite{tang_analysing_2010}. 
However, we will show that no matter how we choose these nodes (e.g., by using a static or temporal metrics to find these path mediators), this strategy is ineffective.
The intuition behind this is given through the example in Figure~\ref{fig:tg2}. Consider the shortest temporal paths from node $A$ to node $F$, namely $(A,C,E,F)$ and a longer (both in terms of hops and time of delivery) temporal path $(A,B,D,E,F)$, also illustrated in Figure~\ref{fig:tgpaths-a}.  If we consider the simple case of patching a single node in an attempt to block the malware from spreading, the best choice would be  node $C$, as the one on most temporal paths, however notice that node $B$ provides an \textit{alternative} path to $F$ albeit a longer path.

Our second strategy relies on the ability to spread a patch message quickly throughout the network; we utilise closeness centrality which is able to capture this property. We now formally describe these temporal centrality measures.
\subsection{Temporal Betweenness Centrality}
The static betweenness centrality of a node $i$ is defined as the fraction of shortest paths between all pairs of nodes which pass through $i$~\cite{wasserman_social_1994}.
Betweenness is commonly used to discover nodes which are critical for information flow and, therefore, prioritised for patching.
Hence, to capture the notion of \textit{temporal} betweenness it is important to take into account not only the number of shortest paths
which pass through a node, but also the \textit{length} of time for which a node
along the shortest path \textit{retains} a malicious message before forwarding
it to the next node. 
For example, consider the 2-hop shortest temporal path from node $A$ to $D$, $(A,B,D)$.  In terms of time, this path could be represented as $(A,B,B,B,D)$ since the malicious message resides on node $B$ for 3 time windows, 
and so we want to assign a higher value as patching this node will lead to a higher probability of stopping the malicious message from spreading.
From this, for a given time window $T$ we define the \textit{temporal betweenness centrality} of node $i$ as:
\begin{equation}
\label{eq:bt}
{\cal B}_i(T)=\frac{1}{(N-1)(N-2)} \sum _{\substack{j \in V \\ j \neq i} }
\sum _{ \substack{k \in V \\ k \neq i \\ k \neq j }}
\dfrac{U(i,T,j,k)}{|\sigma_{j,k}(i)|}
\end{equation}
where the function $U$ returns the number of shortest temporal paths from $j$ to $k$ in which node $i$ has either received a message at time window $T$ or is holding a message from a past time window until the next node is met at some time $T' > T$ and $\sigma_{j,k}(i) \subseteq S_{jk}$ is the set of shortest temporal paths from node $i$ to $j$ which pass through node $i$,  
defined when $\sigma_{j,k}(i) \neq \emptyset$.
In the case when $\sigma_{j,k}(i) = \emptyset$, i.e., node $i$ is totally isolated, we set its betweenness to zero.
Finally, the average temporal betweenness value across all time windows for each node $i$ is:
\begin{equation}
\label{eq:b}
{\cal B}_i=\frac{1}{T} \sum _{t=0}^{T-1} {\cal B}_i(t).
\end{equation}

\subsection{Temporal Closeness Centrality}

Two nodes of a static graph are said to be \textit{close} to each other
if their geodesic distance is small. In a static graph an estimation
of the global \textit{closeness} of a node $i$ is obtained as the
average static shortest path length to all other nodes in the graph~\cite{wasserman_social_1994}.
Similarly, we can extend the definition of closeness to temporal
graphs using the temporal shortest path length between nodes, which is
a measure of how fast a source node can deliver a message to all the
other nodes of the network.  
This can be thought of as a \textit{temporal} minimum spanning tree 
(see Figure \ref{fig:tgpaths-b}).
Given the shortest temporal distance $d_{ij}(t_{min},t_{max})$, 
\textit{temporal closeness centrality} can then be expressed as:
\begin{equation}  
\label{eq:closeness}
{\cal C}_{i}(t_{min},t_{max}) = 1-\left( \frac{1}{W(N-1)} \sum_{j \neq i \in V} d_{i,j}(t_{min},t_{max}) \right)
\end{equation}  
so that nodes that have, \textit{on average}, shorter temporal distances to the other nodes are considered more \textit{central}.
Note that the subtraction from one is only required for a \textit{descending} ranking.
\begin{table}[t!]
\begin{center}
\begin{scriptsize}
\addtolength{\tabcolsep}{-1pt}
\begin{tabular}{|c||c|c|c|}
 \hline
 & CAMBRIDGE  & INFOCOM  & MIT  \\
 \hline
 N  & 18 & 78 & 100 \\
Start Date & 3 Feb 2010 & 23 Apr 2006 & 26 Jul 2004 \\
Duration &  10 Days  & 5 days & 280 days \\
  Scanning Rate     &   30 sec  &  2 min   &  5 min   \\
\hline 
\end{tabular} 
\end{scriptsize}
\end{center}
\vspace{-5pt}
\caption{Experimental Datasets} \label{tbl:datasets}
\vspace{-20pt}
\end{table}
\subsection{Runtime Complexity}
Calculating temporal all pairs shortest path has a time complexity of $O(N^3 T)$.  Since temporal closeness only requires summing across all destination nodes and temporal betweenness only requires an additional summation across all time windows, the asymptotic complexity is the same.

\subsection{Designing a Time-aware Containment Scheme}
We now discuss the potential alternative designs of time-aware containment schemes which utilise temporal centrality measures to find the best node for patching.

\subsubsection{Exploiting Temporal Betweeness Centrality to Block the Paths of Mobile Malware} 
By definition, temporal betweenness centrality finds nodes which mediate between the most communication channels and, hence, their removal will have the greatest impact on the network overall communication efficiency.  
It follows that the first containment scheme can utilise this information to send a patch to these mediating devices, \textit{blocking} a malicious message from using paths which pass through these devices.
As already mentioned, we will show in Section~\ref{sec:individual} that such a scheme is not effective due to many alternative paths which exist in real human contact traces. 
The presence of these alternative paths is due to social clusters during the day which requires a high number of nodes to be patched in order to stop and contain 
the malware.

\subsubsection{Exploiting Temporal Closeness Centrality to Spread a Competitive Patch} 
An alternative scheme can be based on the selection of the best devices to start \textit{spreading} a patch message; the intuition is that a patch message, if started at the right device(s), can propagate faster than the malicious message. Closeness centrality fits this specification since it ranks nodes by their ability to spread a message quickly to the most nodes.  We will show in Section \ref{sec:competitive} that such a scheme is indeed effective.

\section{Evaluation}
\label{sec:evaluation}

\subsection{Experimental Datasets}
\label{sec:datasets}
To evaluate the time-aware mobile malware containment schemes, three traces of real mobile device contacts carried by humans are used: Bluetooth traces of researchers at the University of Cambridge, Computer Laboratory, as part of an emotion sensing experiment \cite{rachuri_emotionsense_mobile_2010}; 
Bluetooth traces of participants at the 2006 INFOCOM conference \cite{cambridge-haggle-2009-05-29}; 
and campus Bluetooth traces of students and staff at MIT \cite{eagle_reality_2006}.  
We shall refer to these as CAMBRIDGE, INFOCOM, MIT, respectively.  
Table \ref{tbl:datasets} describes the characteristics of each set of traces.  
All three datasets were constructed from mobile device co-location where participants were given Bluetooth enabled mobile devices to carry around.  When two devices come into communication range of the Bluetooth radio, the device logs the colocation with the other device. 
For the CAMBRIDGE dataset, all 10 days are used as part of the evaluation.
For the INFOCOM dataset, since devices were not handed out to participants until late afternoon during the first day, only the last 4 days are used.
For the MIT dataset, we show results for the first two weeks of the Fall semester\footnote{http://web.mit.edu/registrar/www/calendar0405.html} representing a typical fortnight of activity.

\label{sec:competitive}
\begin{figure}[t!]
	\begin{center}
		\includegraphics[scale=0.37, trim = 5mm 0mm 0mm 1mm, clip ]{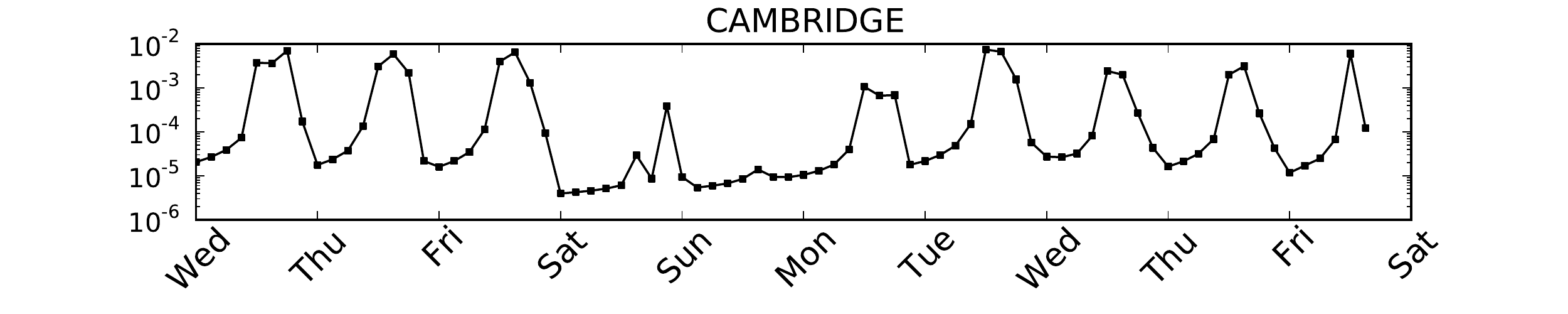}
		\includegraphics[scale=0.37, trim = 5mm 0mm 0mm 1mm, clip ]{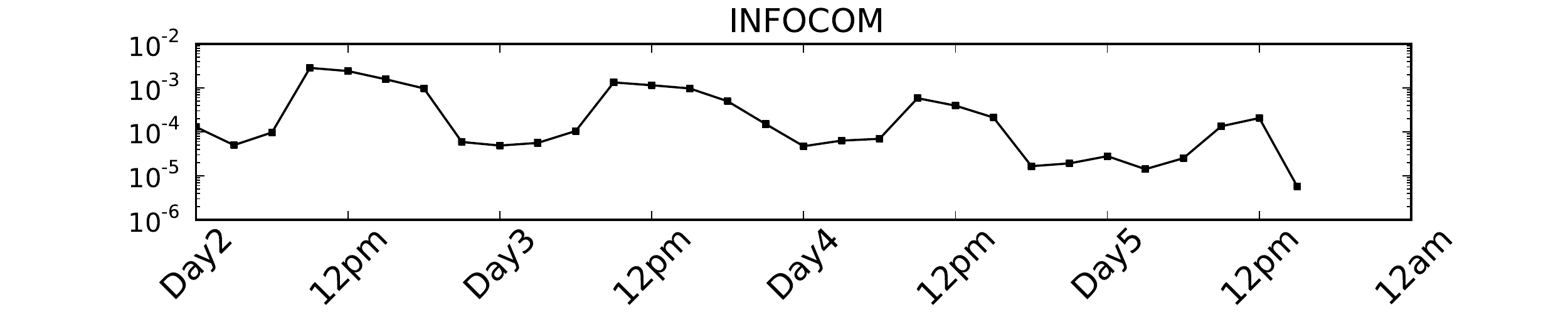}
		\includegraphics[scale=0.37, trim = 5mm 0mm 0mm 1mm, clip ]{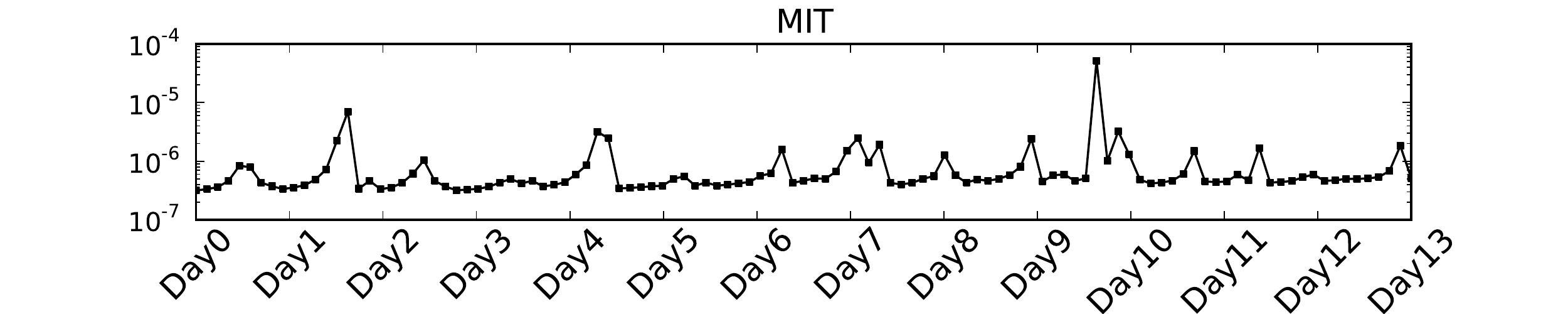}	
	  	\caption{Temporal efficiency (y-axis) as a function of time (x-axis).  Note the logarithmic y-axis.}  \label{fig:eff}
	\end{center}
\vspace{-19pt}
\end{figure}
\subsection{Simulation Setup}
We evaluate the design space of a time-aware containment scheme through a \textit{trace-driven} simulation using as input the three datasets described above. 
We will examine the effects of four key factors: the starting time of the malware spreading process $t_m$ and of the corresponding patching time $t_p$, the initial number of the infected nodes $N_m$ and the initial number of patched nodes $N_p$. 
The top $N_p$ devices are chosen according to the calculated temporal betweenness or temporal closeness centrality ranking from the temporal graph ${\cal G}^w(t_p,t_{max})$, where $w$ is set to the finest window granularity, corresponding to the scanning rate of the devices in each dataset (e.g., 30 second windows for CAMBRIDGE). 
The $N_m$ nodes that are initially infected with malicious messages are chosen uniformly randomly. The results are obtained by averaging over 100 runs for each $N_p$.
The static centralities from the static aggregated graph over the time interval $[t_p,t_{max}]$ are also calculated for comparison.

Our evaluation is based on the following assumptions: firstly, when a node receives a patch message, it is immunised for the rest of the simulation (i.e., we assume that the malware does not mutate over time); secondly, there is always a successful file transfer between devices (errors in transmission can be taken into consideration in the assessment of the contention scheme without changing significantly the results of our work, assuming random transmission failures); thirdly, an attacker chooses nodes at random; and finally, we have no knowledge of which devices are compromised (otherwise the best scheme is to patch those devices immediately).
\begin{figure}[t!]
\vspace{-5pt}
	\begin{minipage}[h!]{0.49\linewidth} 
		\centering
		\includegraphics[scale=0.16, trim = 0mm 0mm 0mm 15mm, clip]{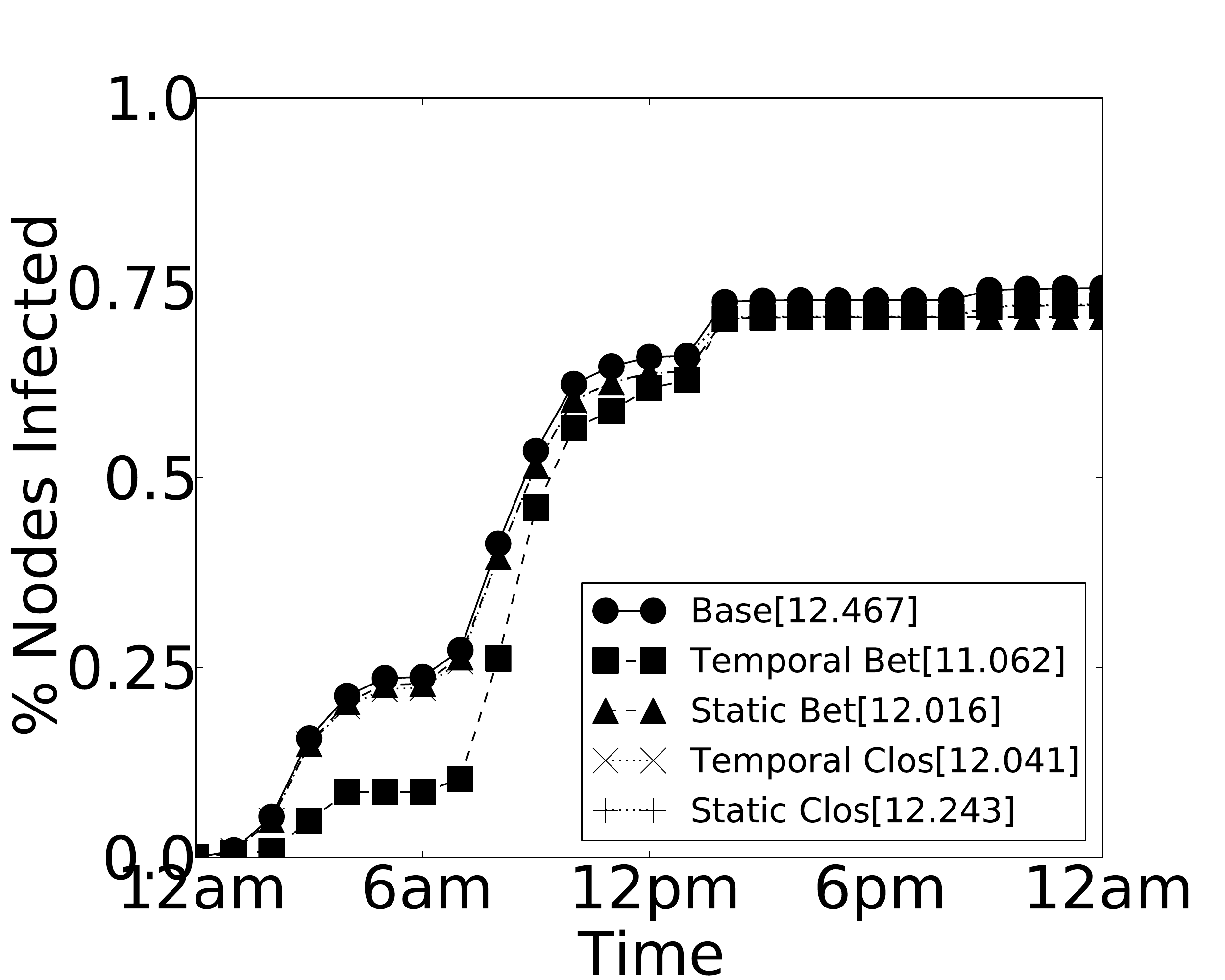}
	\end{minipage}
	\begin{minipage}[h!]{0.49\linewidth}
		\centering
		\includegraphics[scale=0.16, trim = 0mm 0mm 0mm 15mm, clip]{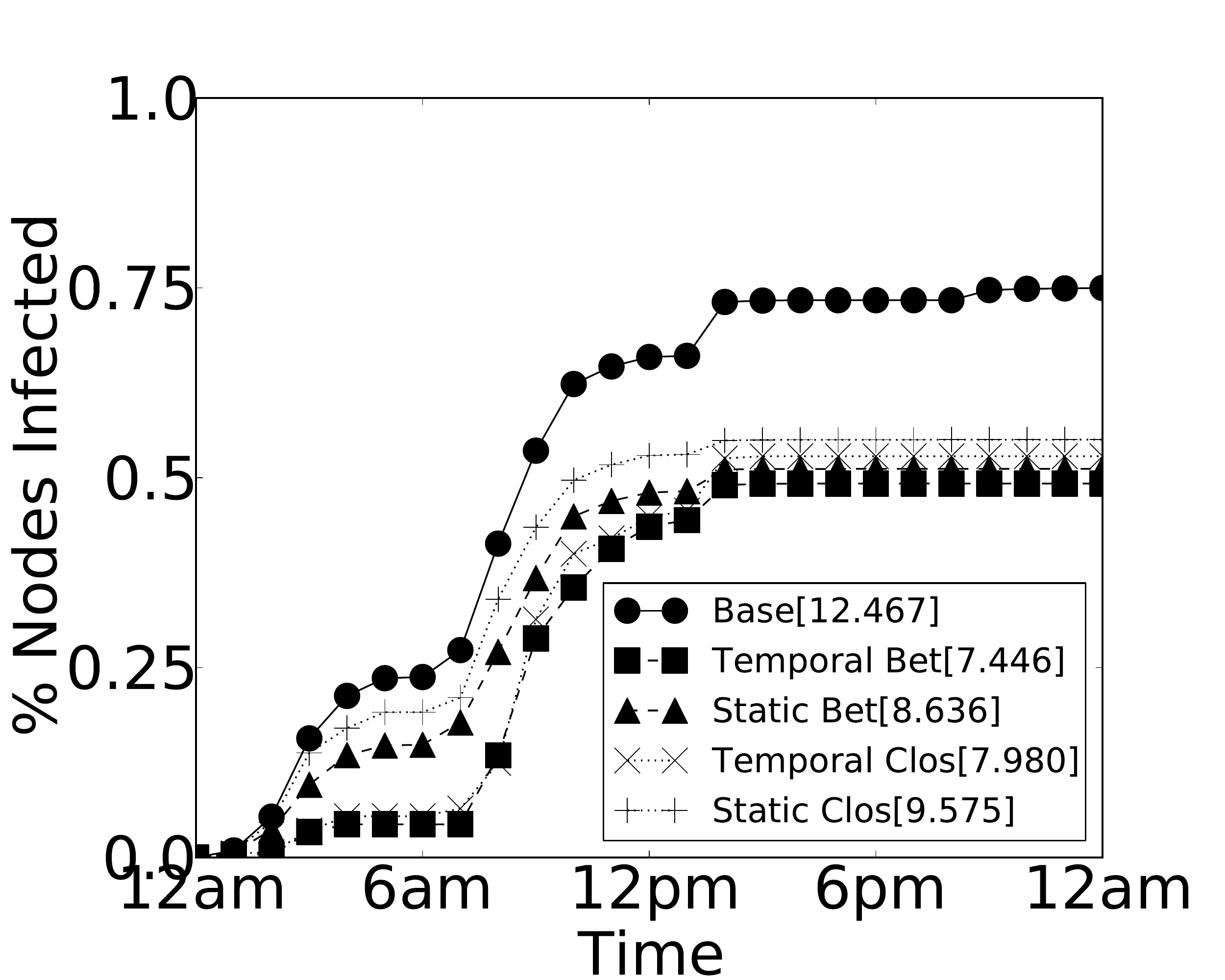}
	\end{minipage}
	\begin{minipage}[h!]{0.49\linewidth} 
		\centering
		\includegraphics[scale=0.16, trim = 0mm 0mm 0mm 10mm, clip]{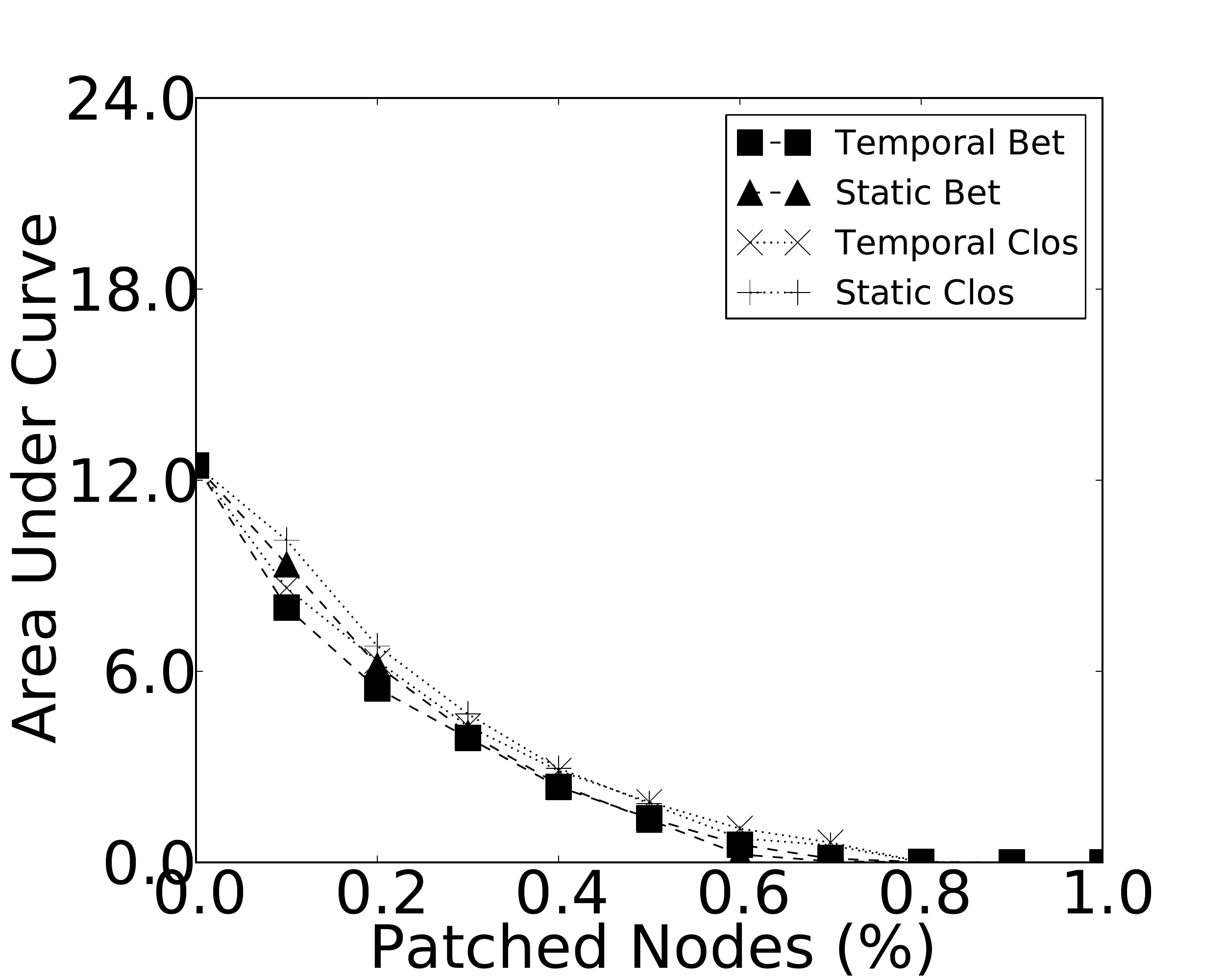}
	\end{minipage}
	\begin{minipage}[h!]{0.49\linewidth}
		\centering
		\includegraphics[scale=0.16, trim = 0mm 0mm 0mm 10mm, clip]{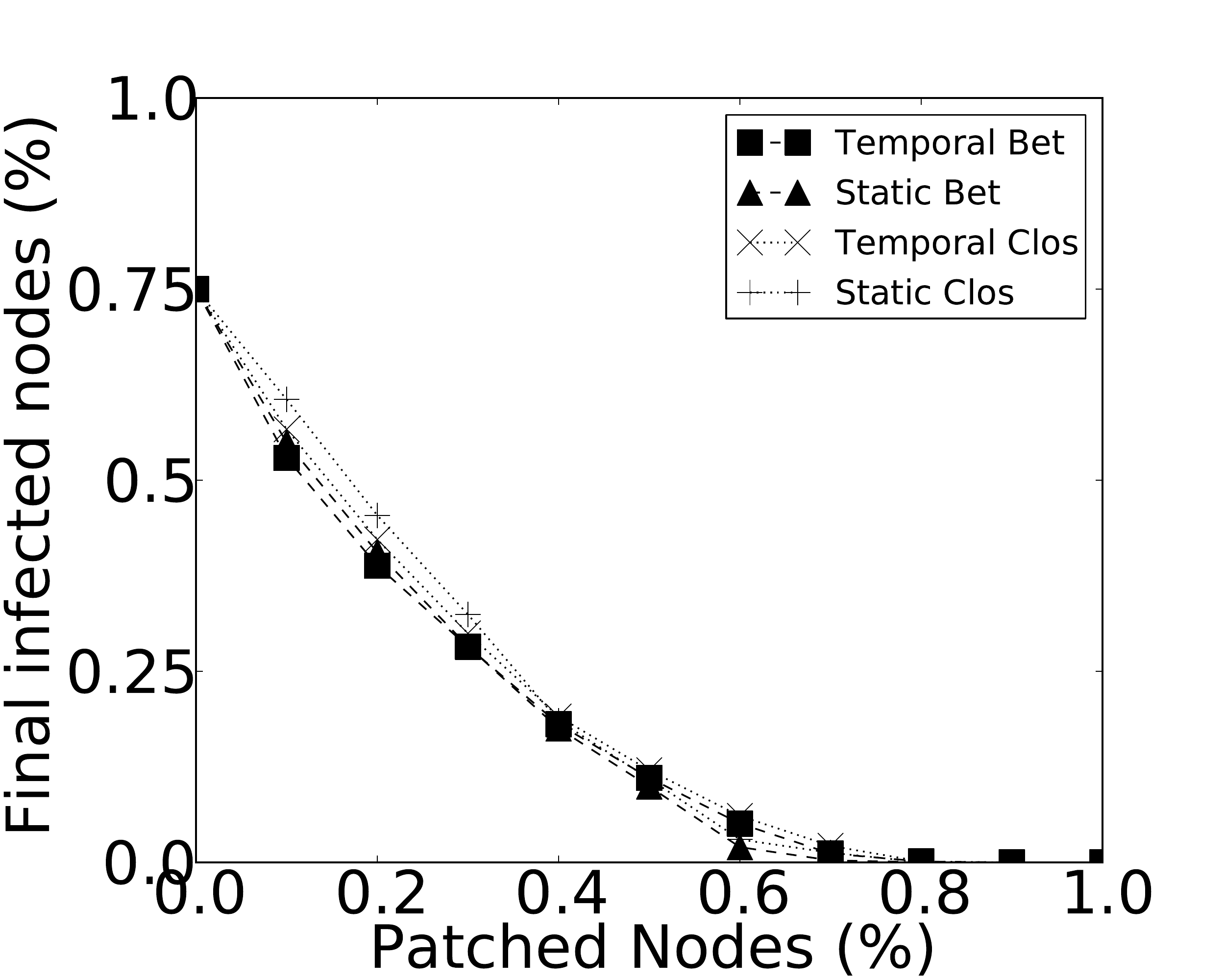}
	\end{minipage}
	\caption{INFOCOM day 4: Immunising 1 (top left) \& 10 source nodes (top right). Area under curves shown in the legend.  Area (bottom left) and final \% of infected nodes (bottom right), as we increase the \% of nodes immunised (x-axis).}
	\label{fig:individual_nodes}
	\vspace{-15pt}
\end{figure}
\subsection{Effects of Time on Malware Spreading}
\label{sec:eff}

Firstly, we briefly analyse the effects of the time of day have on mobile malware propagation.  Let us consider Figure~\ref{fig:eff} where we measure the temporal efficiency (Formula~\ref{eq:Eglob}) as a function of time.
This \textit{sliding} temporal efficiency is calculated for all three datasets.  As we can see there are oscillations corresponding to the natural human periodic daily and weekly behaviour.
For example, the CAMBRIDGE dataset is spread over 10 days, and it is apparent from the traces that a (malicious) message can spread more efficiently during the daytime, as opposed to evenings and weekends.  
\subsection{Non-Effectiveness of Betweenness based Patching}
\begin{figure}[t!]
\vspace{-2pt}
	\begin{center}
		\includegraphics[scale=0.15, trim = 3cm 21mm 35mm 40mm, clip ]{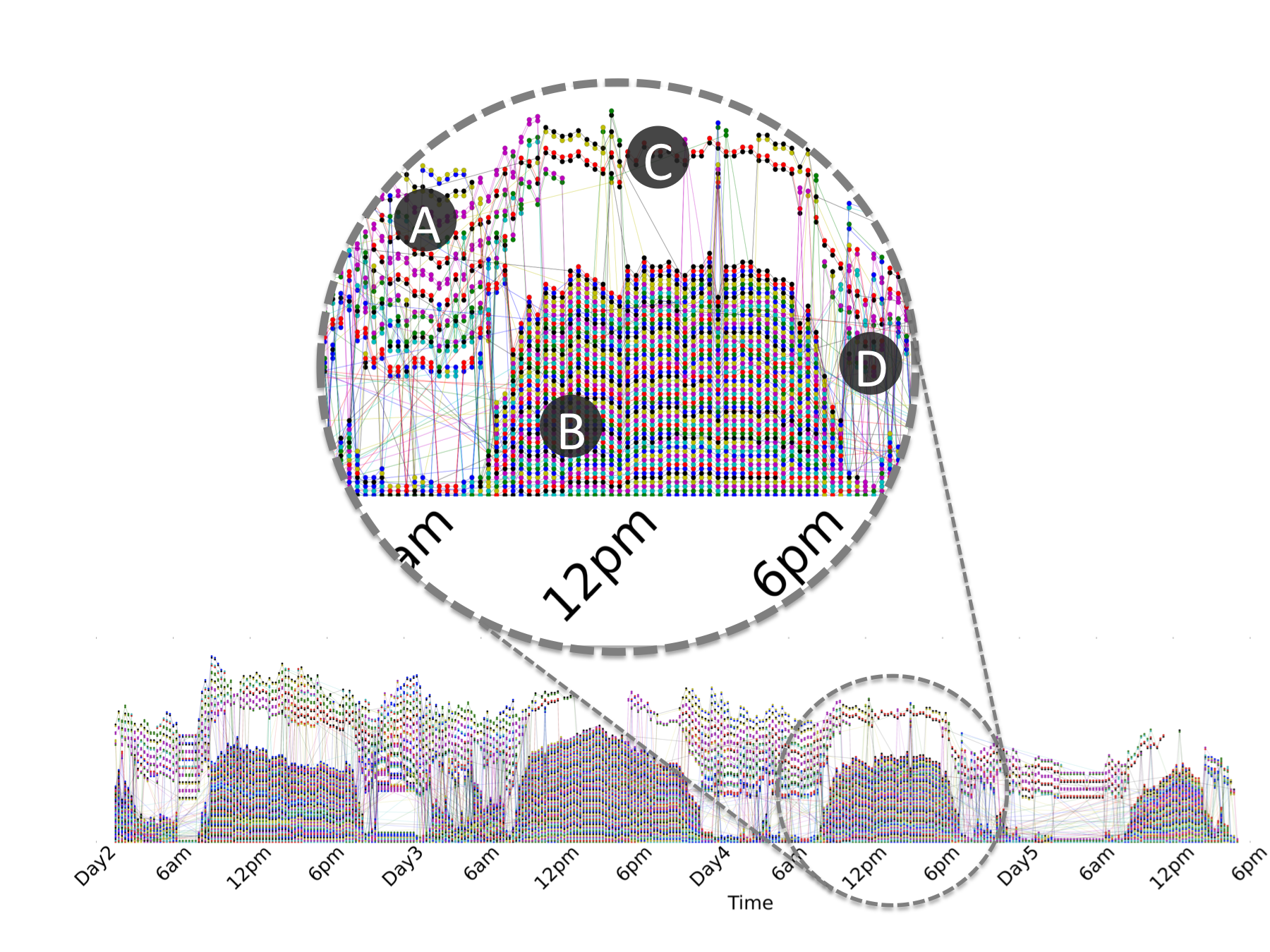}
\vspace{-17pt}
	  	\caption{INFOCOM: Temporal clustering provide four types of alternative paths: (A) inflowing paths to temporal cluster; (B) redundant nodes in cluster; (C) alternative flows around temporal cluster; (D) many outflows to next temporal cluster.}  \label{fig:temporal_clustering}
	\end{center}
\vspace{-21pt}
\end{figure}
Starting from the results of the analysis of the effects time of day has on message spreading, we now evaluate the \textit{best case} scenario for the containment scheme based on patching nodes (without spreading the patch) and we show that this is highly inefficient since it requires a very large number of nodes to be patched via the cellular network 
to be effective.

Using Day 4 of the INFOCOM trace for this example, a piece of malware is started at the beginning of the day ($t_m$=12am) and the device(s) are patched at the same time ($t_p$=12am).  
This is the best case scenario for two reasons: first, the temporal graph in the morning is characterised by low temporal efficiency since there are very few contacts, therefore, the malware spreads slowly (as we have seen in Figure~\ref{fig:eff}); secondly, devices that are immunised immediately have the best chance of blocking malware spreading routes.

Figure~\ref{fig:individual_nodes} shows the ratio of compromised devices across time when the top 1 (top left panel) and top 10 (top right panel) devices are patched after being selected using betweenness and closeness.
As we can see, temporal betweenness 
initially performs better than static betweenness and both temporal and static closeness (quantified by the difference in the area under each curve, shown in the legend). However, by 7am we observe a steep rise in the number of compromised devices and by the end of the day, all curves converge to the same point.  We also note that {\em in both cases it is not possible to totally contain the malware, suggesting that more devices need to be patched}.  Taking a broader view, Figure~\ref{fig:individual_nodes} shows the area under the curve (bottom left) and final ratio of nodes infected (bottom right) as we increase the number of patched devices.  
Clearly, even when the malware is started at the slowest time of day for communication, we still need to patch 80\% of the devices before we can completely stop the malware from spreading; this can be considered an impractically high number of devices to patch.  Similar high percentages are also required in the MIT trace with a minimum of 45\% patched nodes.
We can also conclude that in human contact networks, even with blocked nodes, it is only a matter of time before a (malicious) message disseminates to all nodes.
To understand the reason for the effectiveness of a (malicious) message propagation, we take a visual analysis approach: Figure \ref{fig:temporal_clustering} shows the \textit{temporal activity diagram}\footnote{This plot was inspired by http://xkcd.com/657} for the INFOCOM experiment across all four days.  
This gives a bird's eye view of proximity between individuals as they move between groups of colocated people across time, where the trajectory of the same node is given by a straight line.  
The horizontal axis is time and the vertical groupings of nodes represents people that are in the same
static connected component such that there is a path between every node in that cluster.
The main feature to note is the \textit{temporal cluster} of remarkable size which appears from around 7am until 7pm every day, coinciding with the main activities at the INFOCOM conference\footnote{http://www.ieee-infocom.org/2006/technical\_program.htm}.  
 By means of this infographic, what we see are periodic clusters of nodes during the daytime and smaller disparate clusters during the evening.  
Figure~\ref{fig:temporal_clustering} also zooms into Day 4, highlighting the four types of activity which give rise to temporal clustering and more importantly, to alternative paths providing link redundancy for a message to pass through a network over time.
{\em Since this strategy cannot deal with these alternative paths effectively, the propagation of a malicious message can merely be \textit{slowed} down}.
Hence, the rapid increase of infected nodes that can be observed in Figure~\ref{fig:individual_nodes} around 7am can be attributed to the presence of this large  temporal cluster starting at 7am where many alternative paths are present and, therefore, the spreading cannot be stopped just patching some of the nodes.
We conclude that this containment strategy is not efficient given the large number of patch messages it requires.

\label{sec:individual}
\subsection{Effectiveness of Closeness based Patching (Worst Case Scenario)}
\label{sec:competitive}
Since the blocking based containment scheme is not effective, we now evaluate the closeness based \textit{spreading} scheme with the aim of disseminating a patch message throughout the network more quickly than a malicious message.   
\begin{figure}[t!]
	\begin{minipage}[h!]{0.49\linewidth} 
		\centering
		\includegraphics[scale=0.28, trim = 0mm 0mm 0mm 3mm, clip]{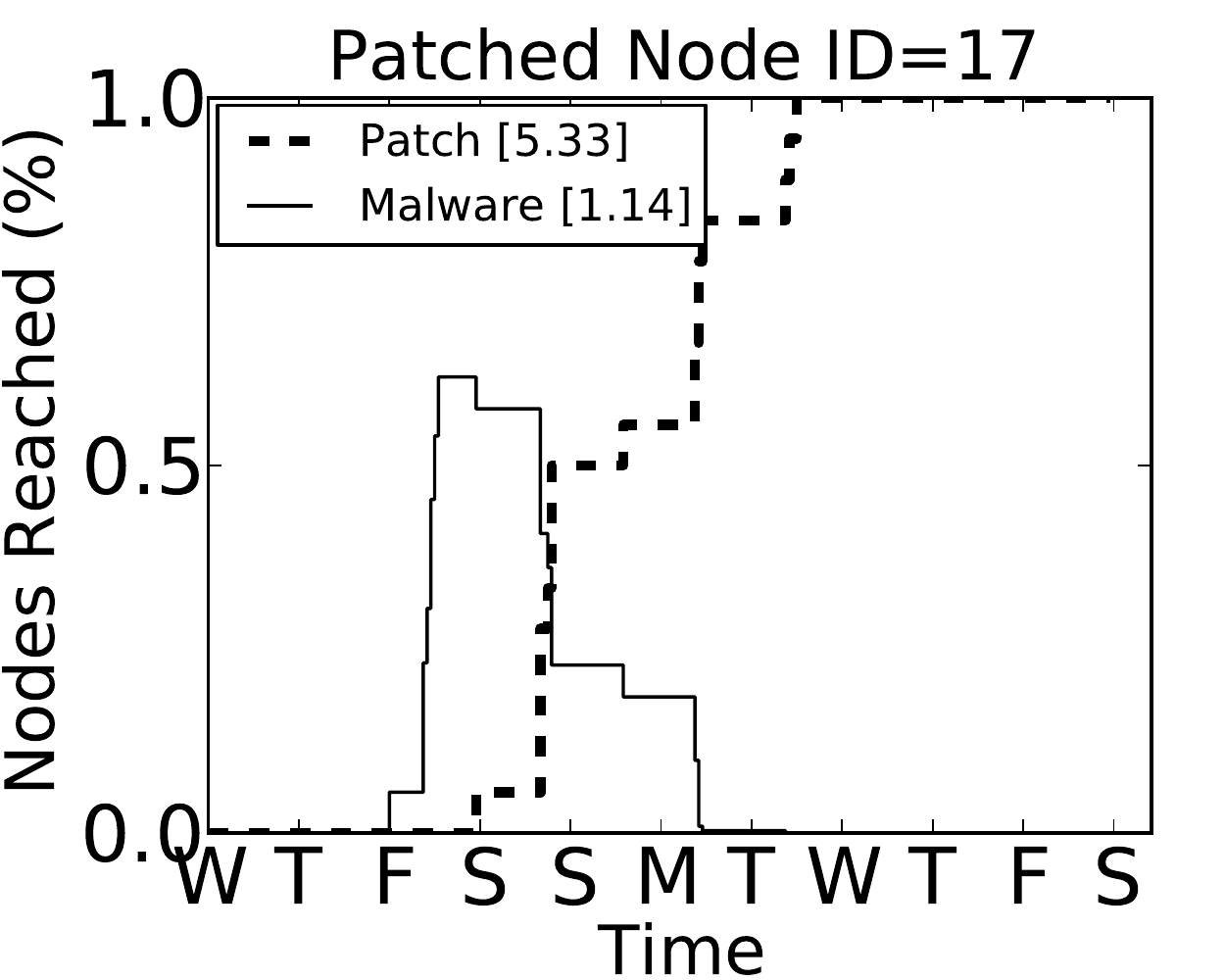}
	\end{minipage}
	\begin{minipage}[h!]{0.49\linewidth}
		\centering
		\includegraphics[scale=0.28, trim = 9mm 0mm 0mm 3mm, clip]{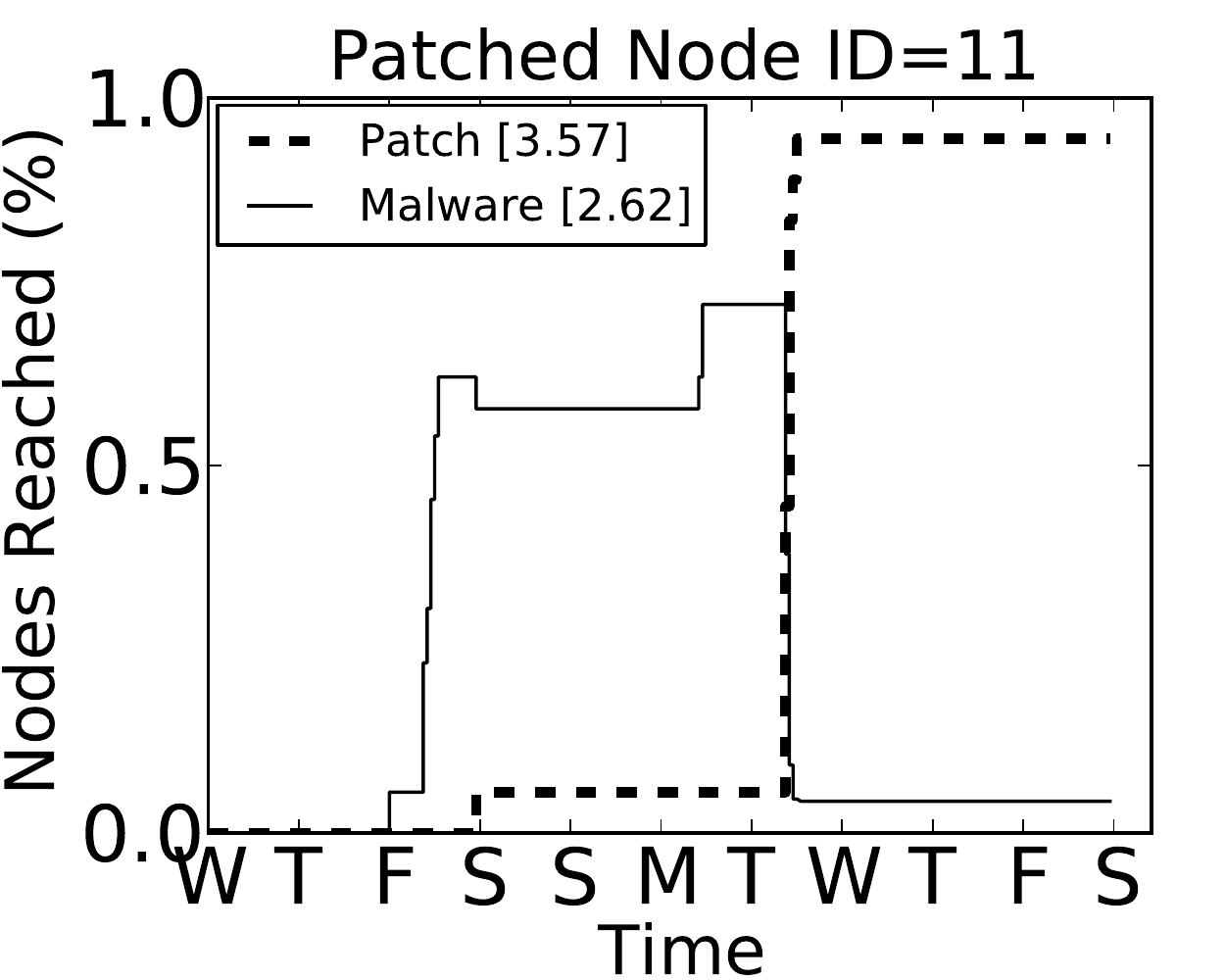}
	\end{minipage}
	\caption{CAMBRIDGE [$t_m$=Fri 12am, $t_p$=Sat 12am] delivery rate (y-axis) starting a mobile worm from single node. 
Best case (left) and worse case patching node (right) shown. Area under curve presented in legend.}\label{fig:ex1}
\vspace{-8pt}
\end{figure}
\begin{figure}[t!]
	\begin{minipage}[h!]{0.49\linewidth} 
		\centering
		\includegraphics[scale=0.13, trim = 0mm 0mm 0mm 8mm, clip]{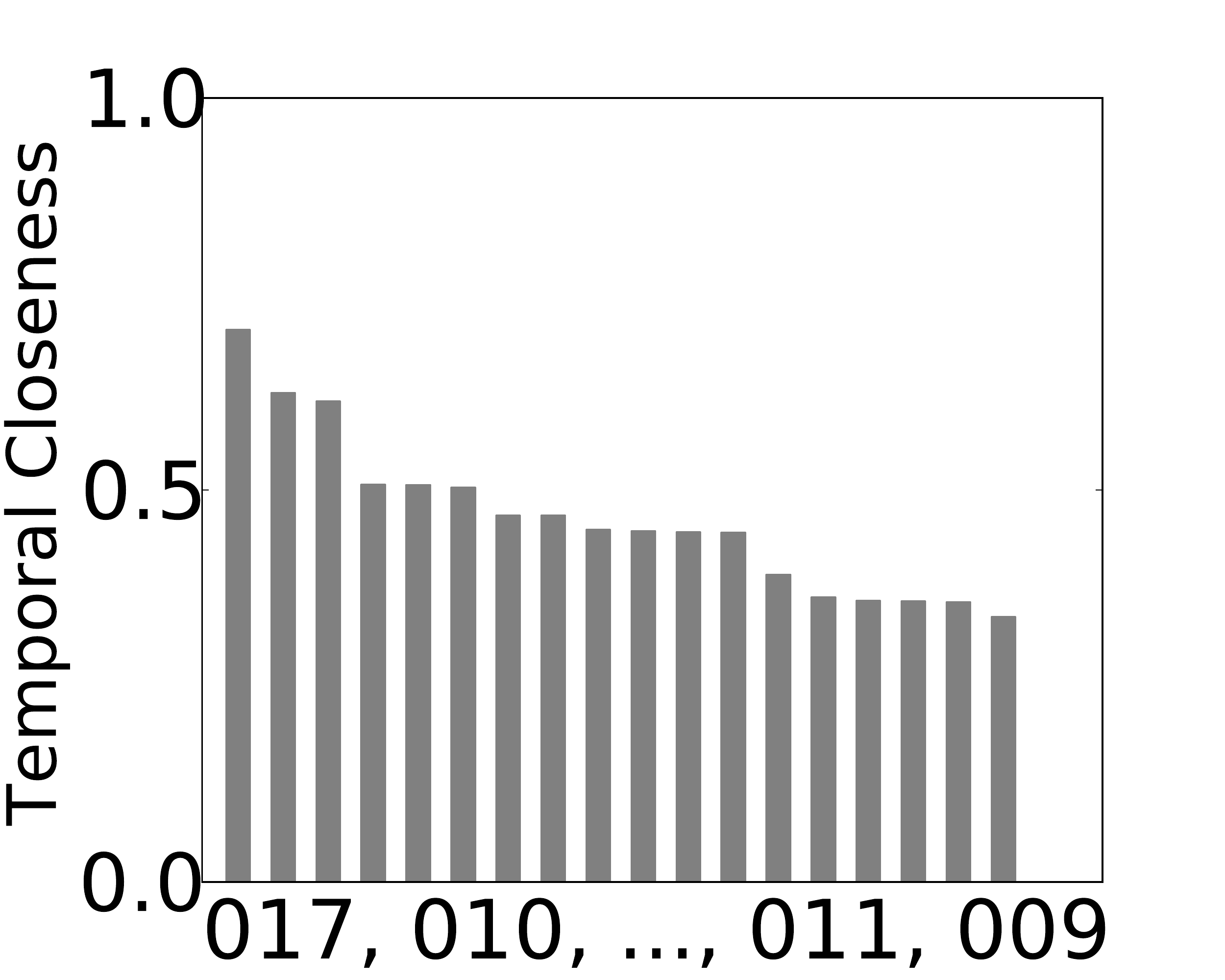}
	\end{minipage}
	\begin{minipage}[h!]{0.49\linewidth}
		\centering
		\includegraphics[scale=0.13, trim = 0mm 0mm 0mm 8mm, clip]{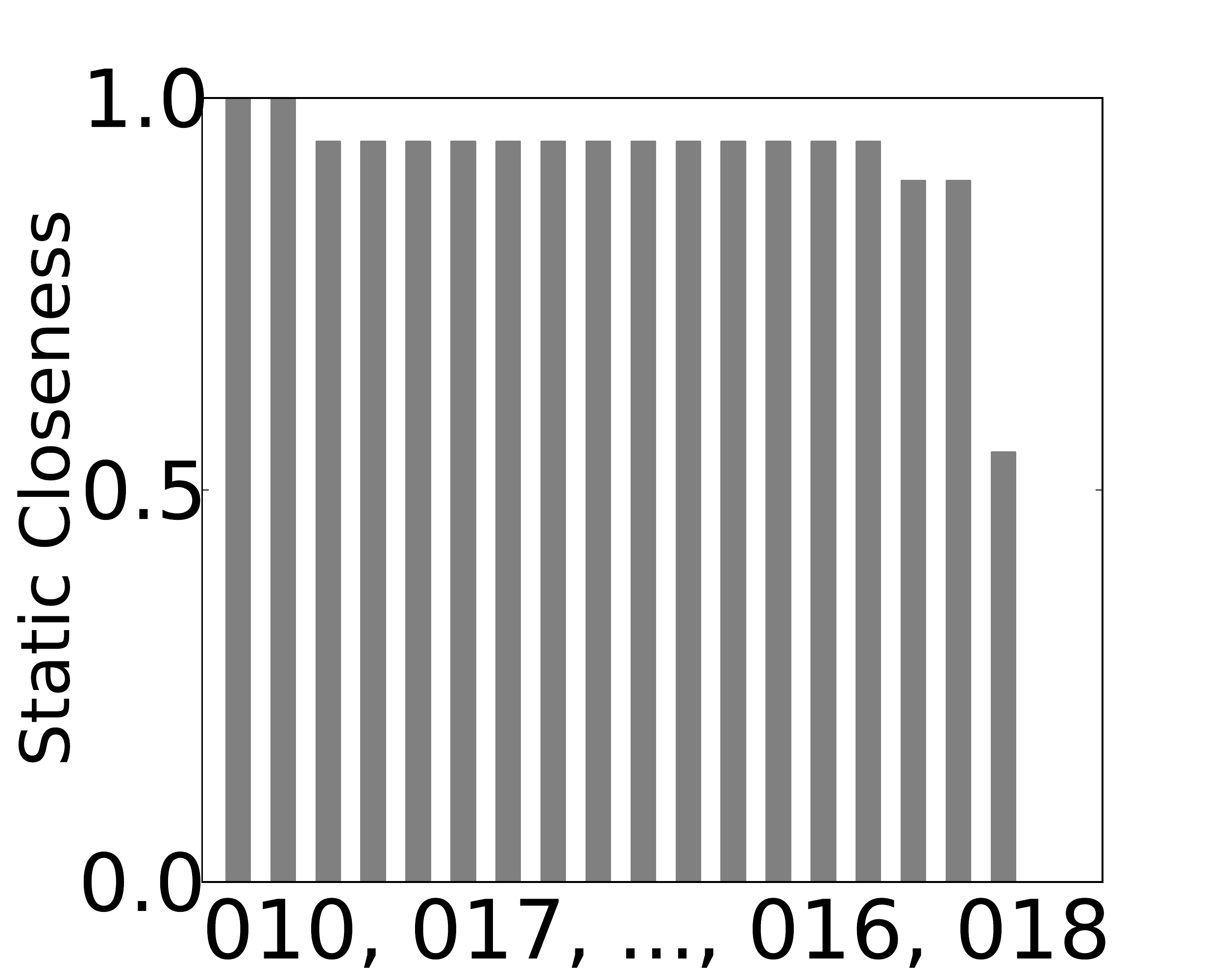}
	\end{minipage}
	\caption{Temporal (left) and static (right) closeness centrality ranking for Figure \ref{fig:ex1}. Top two and bottom two device IDs shown on x-axis.  Nodes ranked left to right.}\label{fig:ex1_centrality}
\vspace{-8pt}
\end{figure}
\begin{figure}[t!]
	\begin{minipage}[h!]{0.49\linewidth} 
		\centering
		\includegraphics[scale=0.3, trim = 16mm 4mm 0mm 0mm, clip]{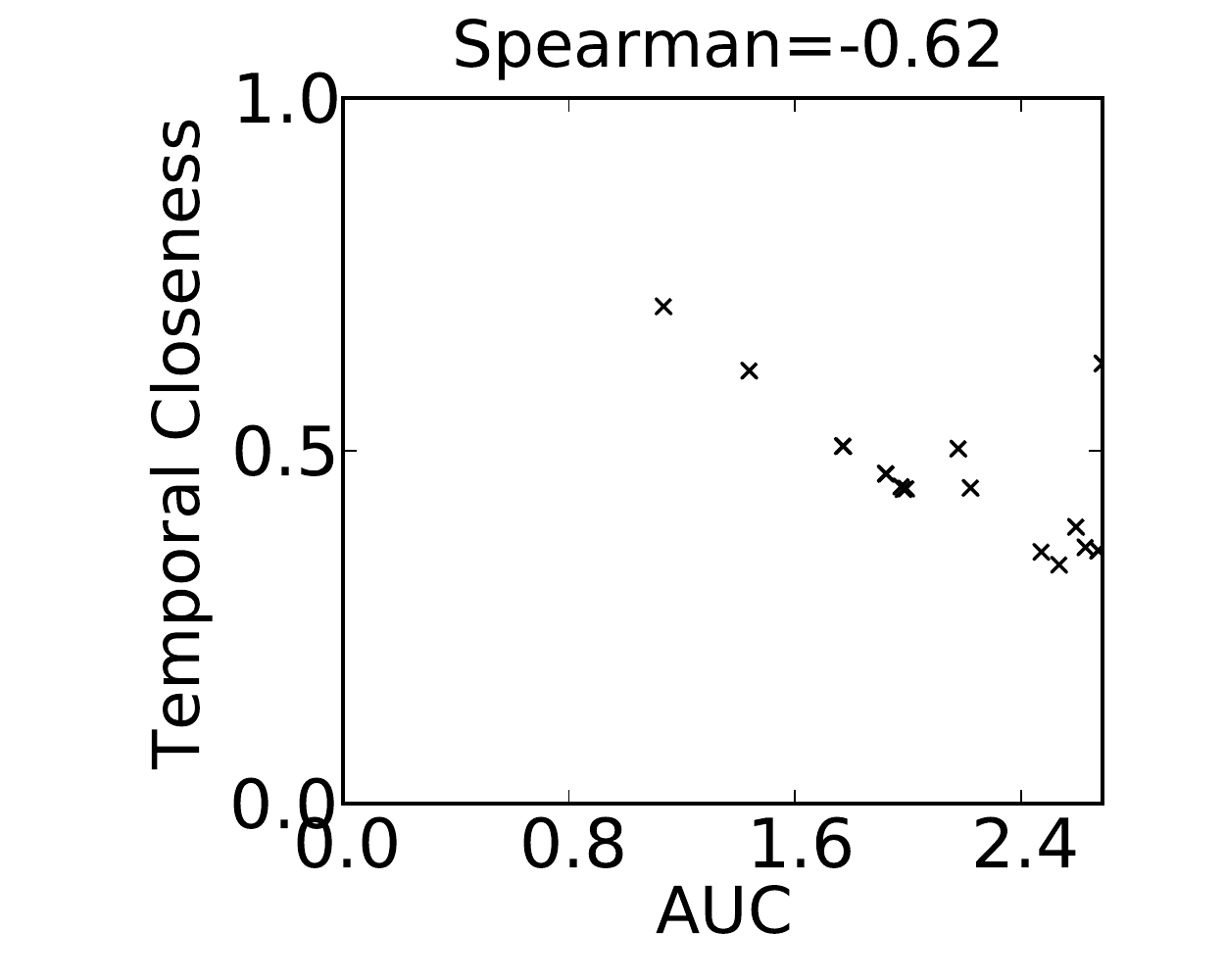}
	\end{minipage}
	\begin{minipage}[h!]{0.49\linewidth}
		\centering
		\includegraphics[scale=0.3, trim = 16mm 4mm 0mm 0mm, clip]{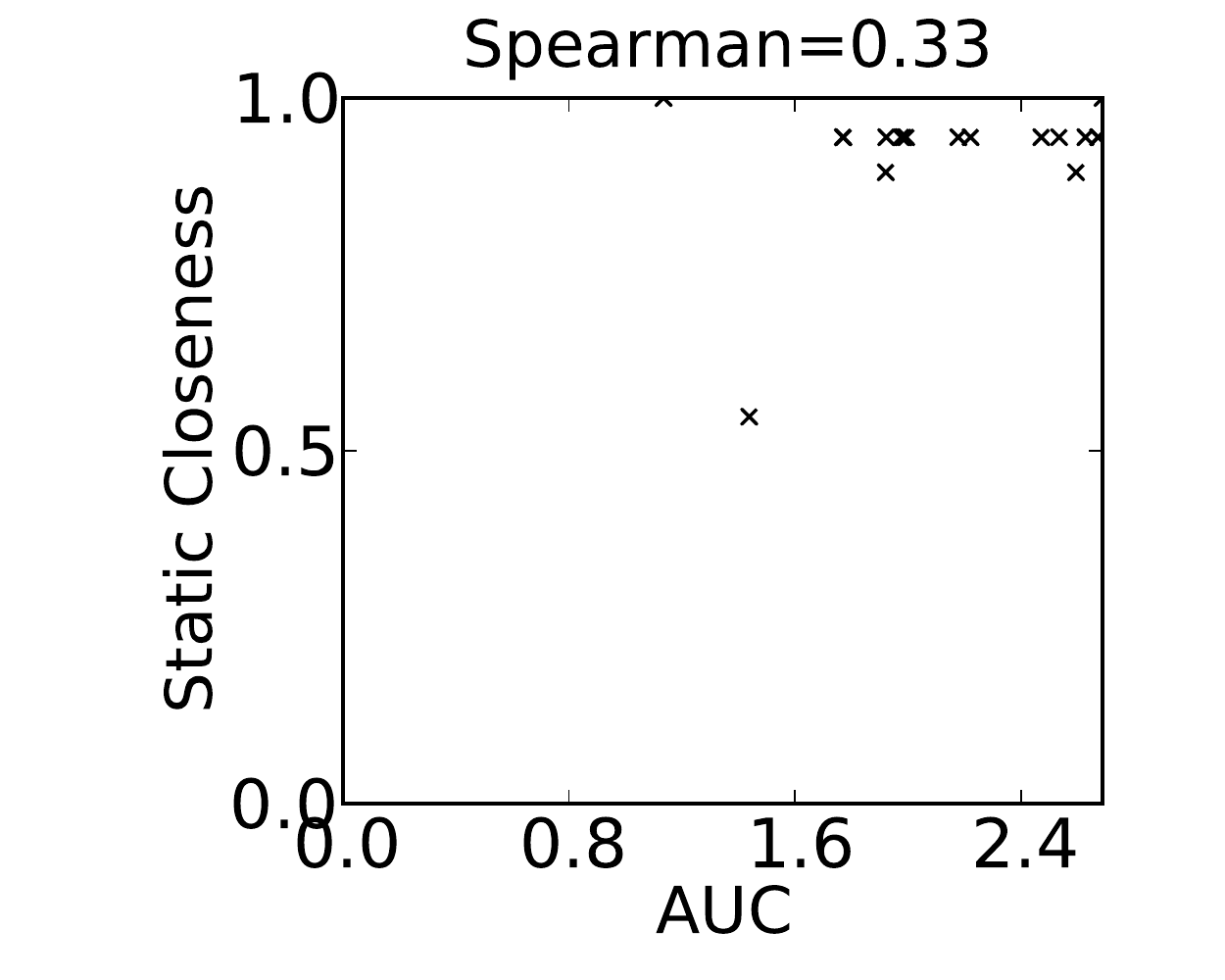}
	\end{minipage}
\vspace{-3pt}
	\caption{Correlation between AUC with temporal (left) and static (right) closeness centrality.}\label{fig:ex1_centrality_corr}
\vspace{-15pt}
\end{figure}
\begin{figure*}[t!]
\vspace{-5pt}
	\begin{minipage}[h!]{0.31\linewidth} 
		\centering
		\includegraphics[scale=0.26, trim = 0mm 6mm 45mm 1mm, clip]{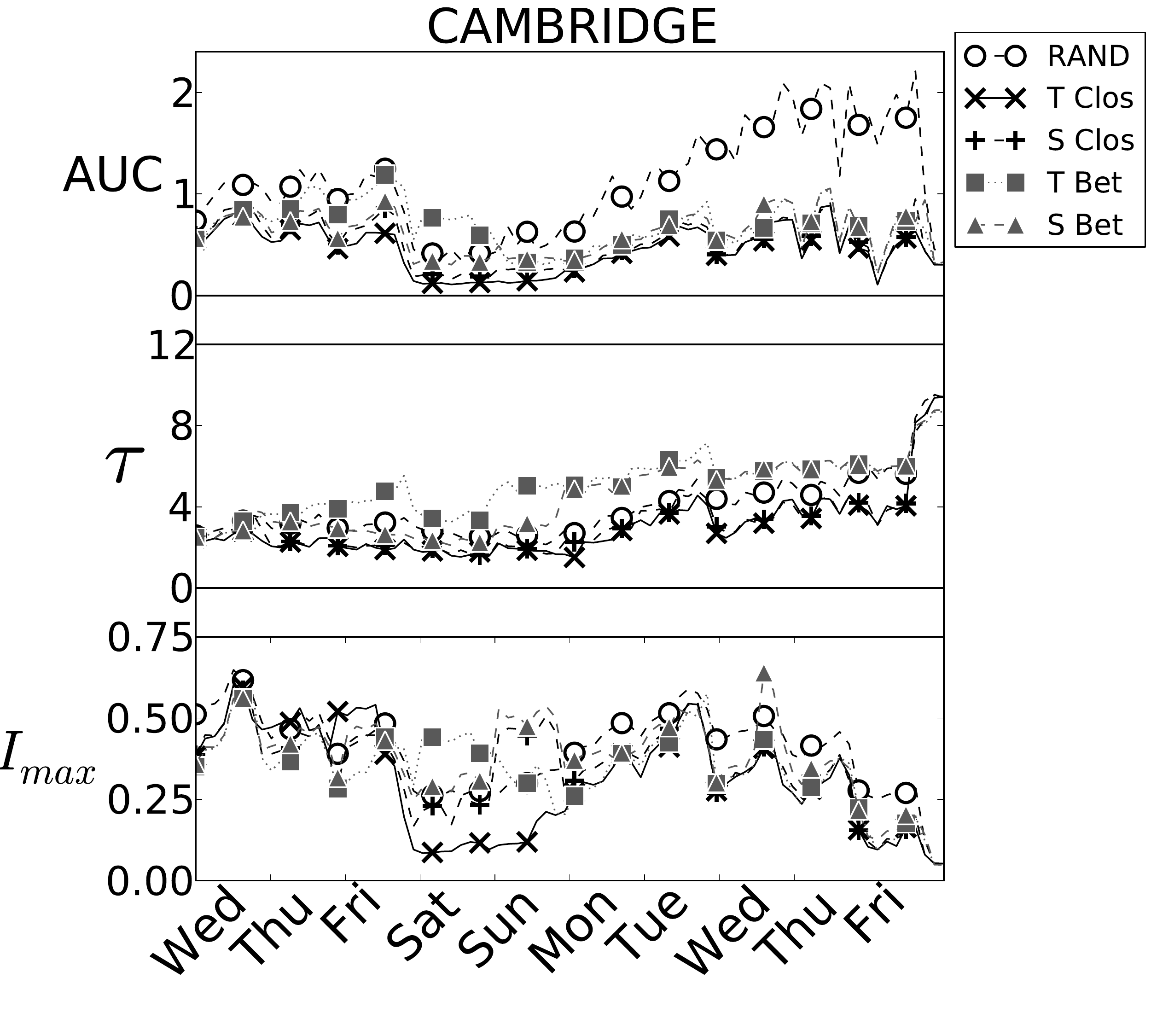}
	\end{minipage}
	\begin{minipage}[h!]{0.31\linewidth}
		\centering
		\includegraphics[scale=0.26, trim = 0mm 6mm 45mm 1mm, clip]{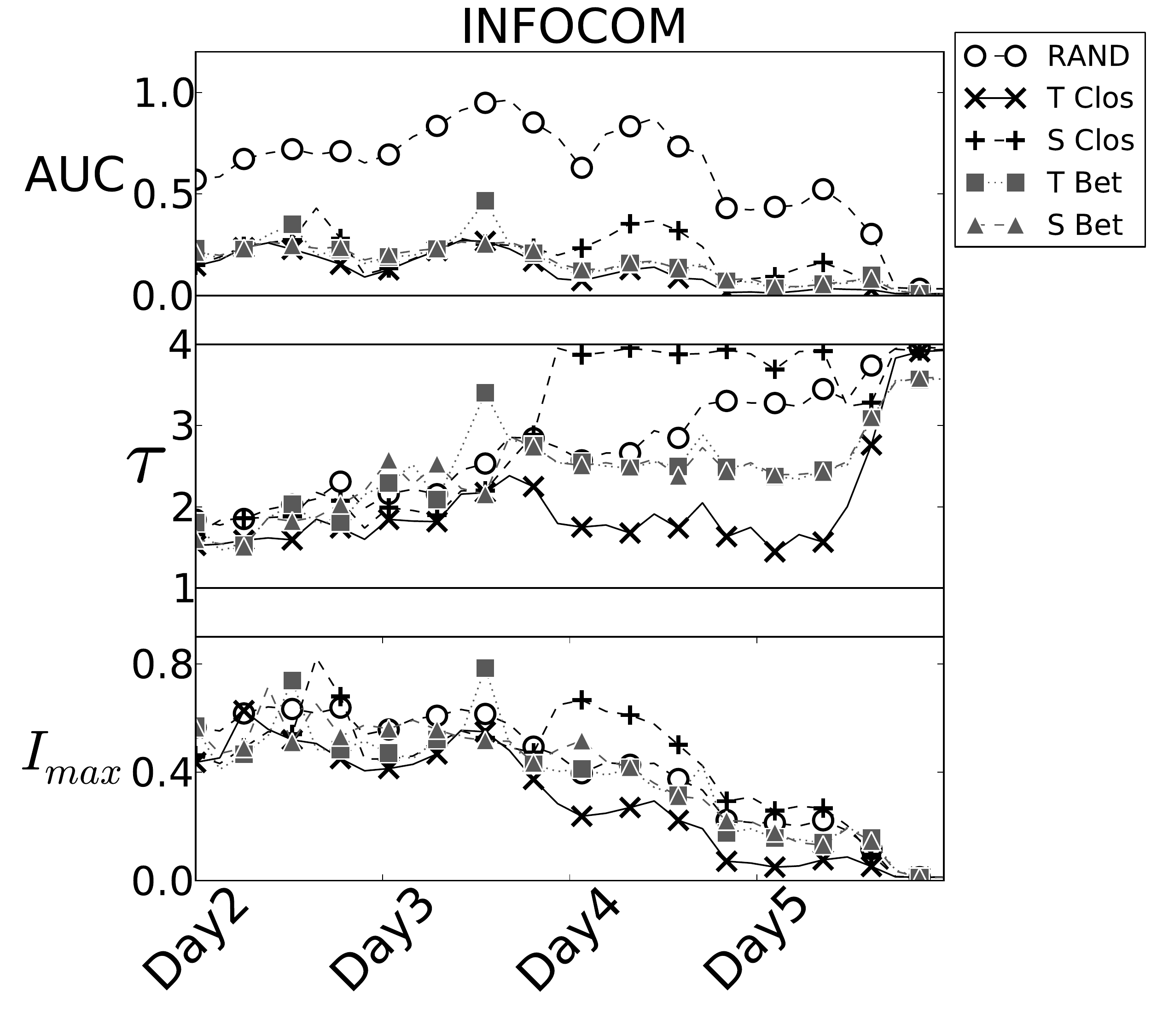}
	\end{minipage}
	\begin{minipage}[h!]{0.35\linewidth}
		\centering
		\includegraphics[scale=0.26, trim = 0mm 6mm 0mm 1mm, clip]{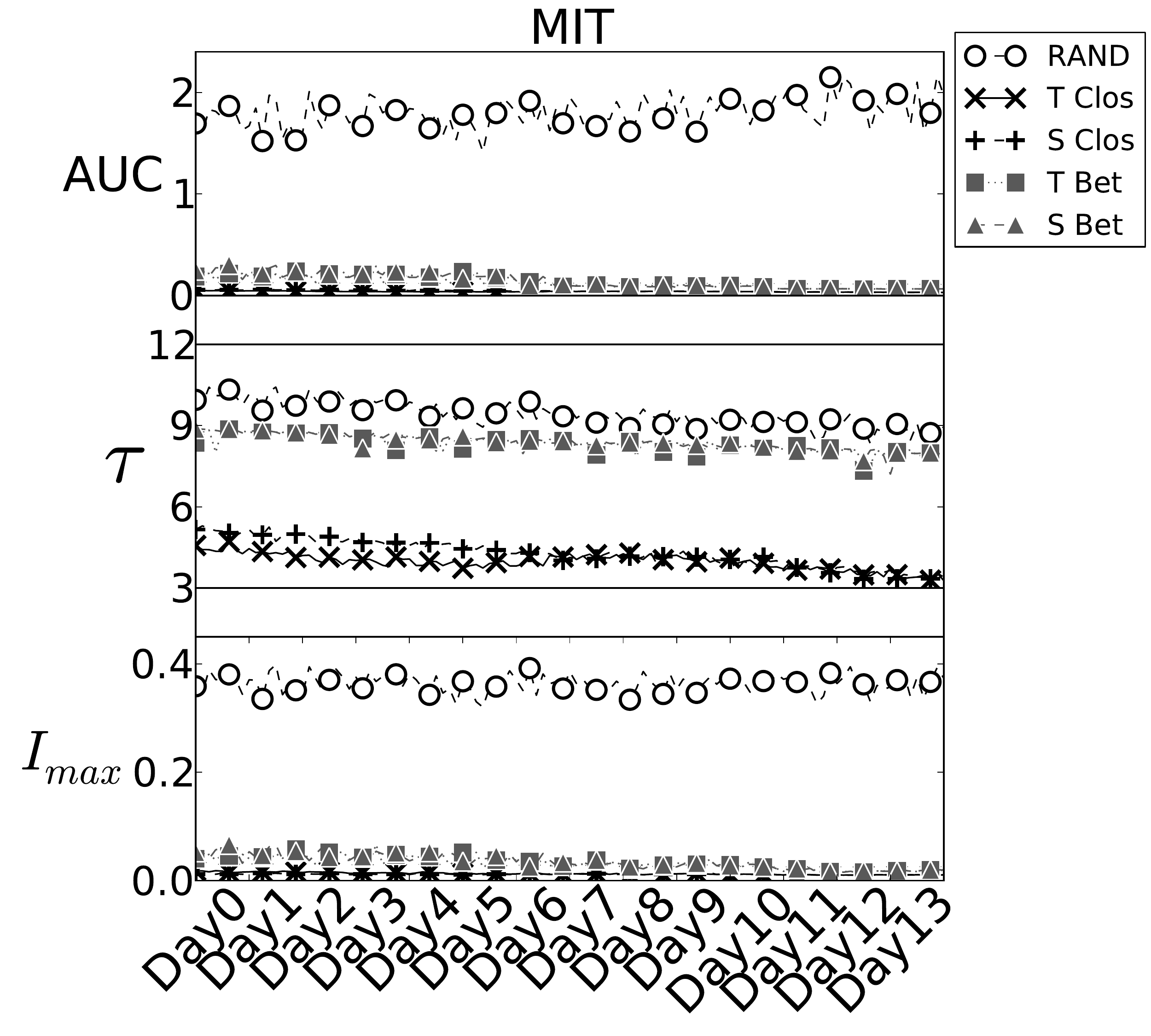}
	\end{minipage}
\vspace{-2pt}	
	\caption{Performance of temporal, static and naive node selection, across different malware start times (x-axis), averaged over all patch delays.}\label{fig:Iall_N1_Pavg}	
	\label{fig:Iall_N1_Pavg}
\vspace{-12pt}
\end{figure*}
We start our analysis by examining a \textit{worst case} scenario using the CAMBRIDGE dataset: a researcher receives a malicious message on their device in the early hours of Friday morning ($t_m$=Fri 12am) and the malicious program replicates itself to any devices it meets during the day. A patch message is started a day later to try patching all the compromised devices ($t_p$=Sat 12am).  Again referring to Figure~\ref{fig:eff}, this can be considered as a worst case since the malware is started during a day with high spreading efficiency and the patch is delayed until the weekend when the efficiency is low.

Figure~\ref{fig:ex1} shows the spreading rate for the malicious message versus the best (left) and worst device (right) to start the patching message. These results were obtained by running simulations considering every single device as a starting point of the patching process, and then ranking them based on three \textit{performance metrics}:
\begin{itemize}
\item the area under the curve (AUC), which captures
the behaviour of the infection over time with respect to the number of infected devices\footnote{The AUC is commonly used in epidemiology and medical trials \cite{trials_small_2001}.};
\item the peak number of compromised devices ($I_{max}$);
\item the time in days necessary to achieve total malware containment ($\tau$).
\end{itemize}
Since  the AUC captures both the $I_{max}$ and $\tau$, the best and worst initial devices that were patched were selected using the AUC.  Comparing all three measures, the case related to the selection of the worst device (right panel) is characterised by double AUC (2.62 vs. 1.07); a higher peak in compromised devices $I_{max}$ (68\% vs. 60\%) and by the fact that it is not possible to fully contain the malware in a finite time $\tau$ ($\infty$ vs. 3.3 days). 
Now comparing these observations with centrality, in Figure~\ref{fig:ex1_centrality} we observe that the node characterised by the highest temporal closeness centrality (ID=17) is also the optimal one for spreading the patch and the node that leads to the worst performance (ID=11) is ranked within the bottom two nodes.
This should be compared with static centrality which ranks the best device to start the patching process (ID=17) in second place and the worst device (ID=11) seventh from the bottom (not shown).
Also, the values of static centrality of each node is more uniformly distributed; a fact which can be attributed to the dense static graph previously observed in Figure~\ref{fig:tg}.  
The stronger correlation between temporal closeness centrality and an effective malware containment scheme can be seen more clearly by plotting these rankings against the AUC in Figure~\ref{fig:ex1_centrality_corr}.
We expect a strong negative correlation since centrality values are ranked in \textit{descending} order;
by using temporal closeness centrality, we can identify the best node to start disseminating a patch message to contain a piece of mobile malware which fits our intuition that spreading a patch message quickly is the best containment strategy.
\subsection{Effects of Temporal Variability}
Thus far we have only considered a single malware start time. We now take a broader view and examine the effects of varying malware start time ($t_m$) and patch delay ($t_p$).  For each dataset the AUC, $I_{max}$ and $\tau$ are exhaustively calculated for different malware start times at hourly intervals
and increasing patch delays starting from zero (i.e., patch messages start at the same time as malicious messages) to up to 2 days.  
We compare node selection based on temporal and static closeness to that of temporal and static betweenness.
As a baseline, a naive method of randomly selecting patching nodes is also calculated, averaged over 100 runs.
\subsubsection{Sensitivity to Malware Start Time}
To understand the effects of a malicious message starting at different times, Figure~\ref{fig:Iall_N1_Pavg} shows for each dataset the performance metrics as a function of the malware start time $t_m$, averaged over all patch delays.  
Firstly, referring back to the temporal efficiency from Figure~\ref{fig:eff}, which exhibited daily peaks and troughs during the weekend, the AUC and the maximum number of infected nodes $I_{max}$ tend to follow these same patterns (strictly related to human circadian rhythms); 
however, the total time of containment ($\tau$) remains stable across all start times.
These results demonstrate that this time-aware containment scheme is an effective method of quickly containing malware, irrespective of when the malware started.
\begin{figure*}[t!]
\vspace{-5pt}
	\begin{minipage}[h!]{0.35\linewidth} 
		\centering
		\includegraphics[scale=0.225, trim = 0mm 11mm 0mm 5mm, clip]{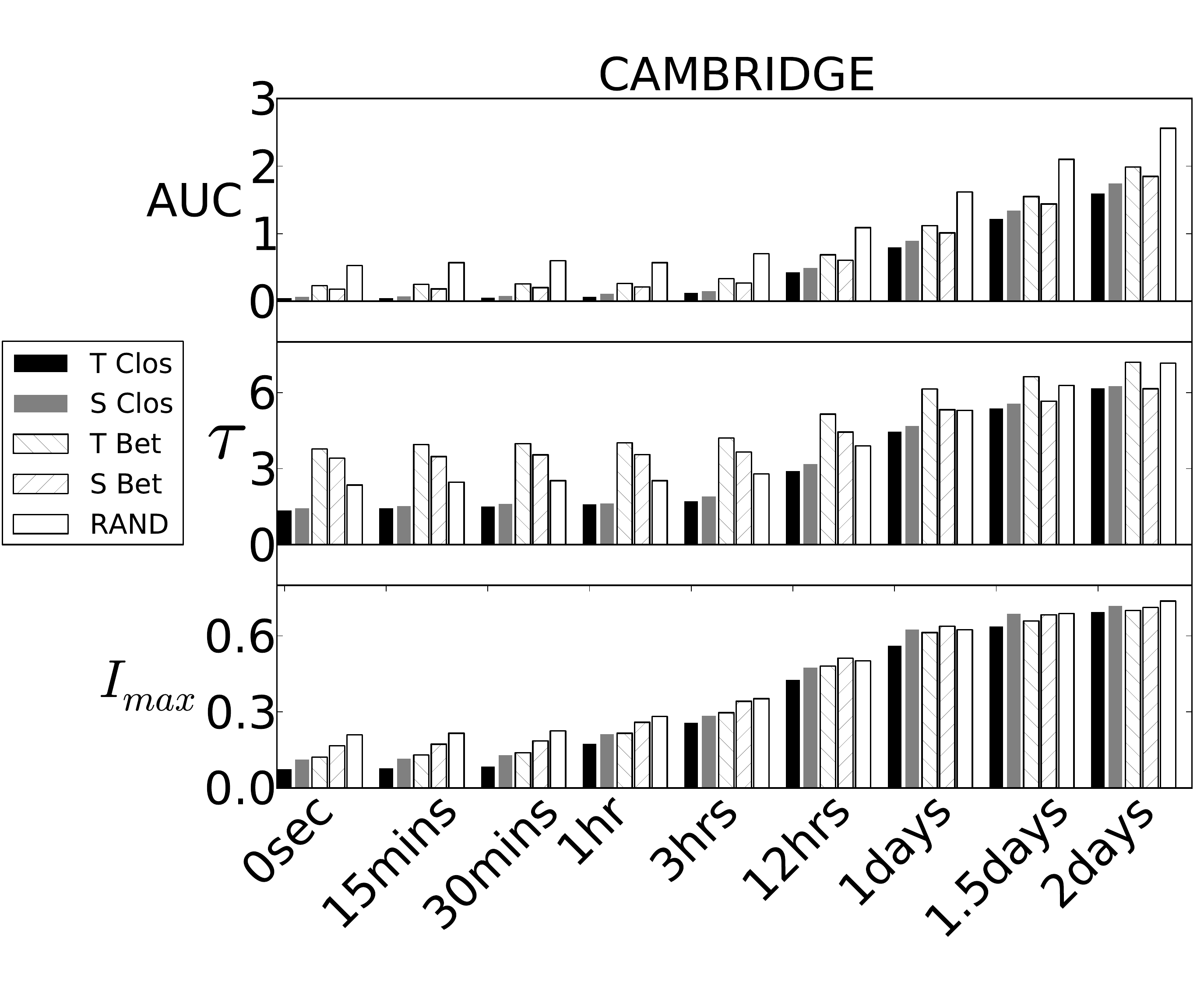}
	\end{minipage}
	\begin{minipage}[h!]{0.32\linewidth}
		\centering
		\includegraphics[scale=0.225, trim = 10mm 11mm 0mm 5mm, clip]{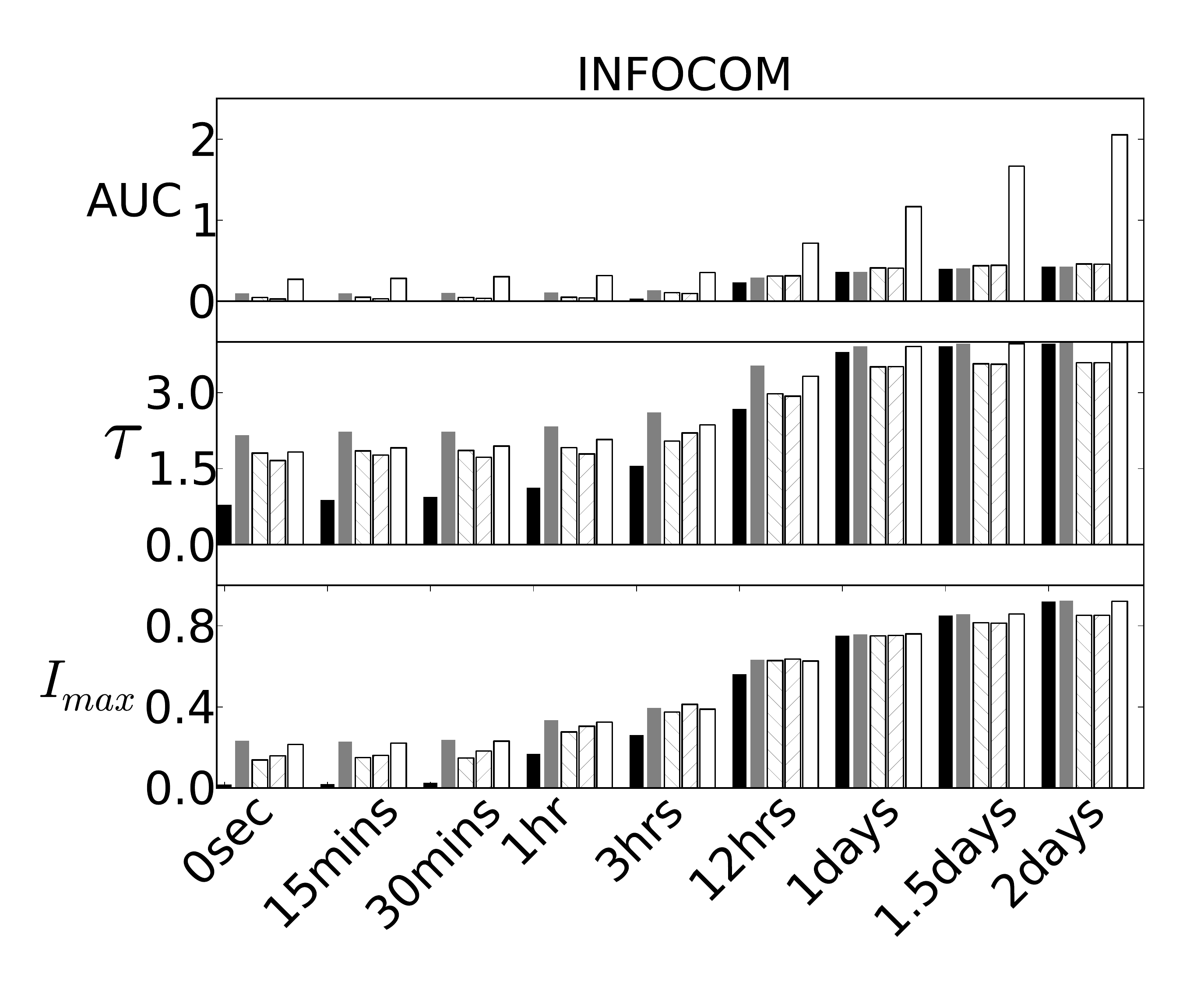}
	\end{minipage}
	\begin{minipage}[h!]{0.32\linewidth}
		\centering
		\includegraphics[scale=0.225, trim = 8mm 11mm 10mm 5mm, clip]{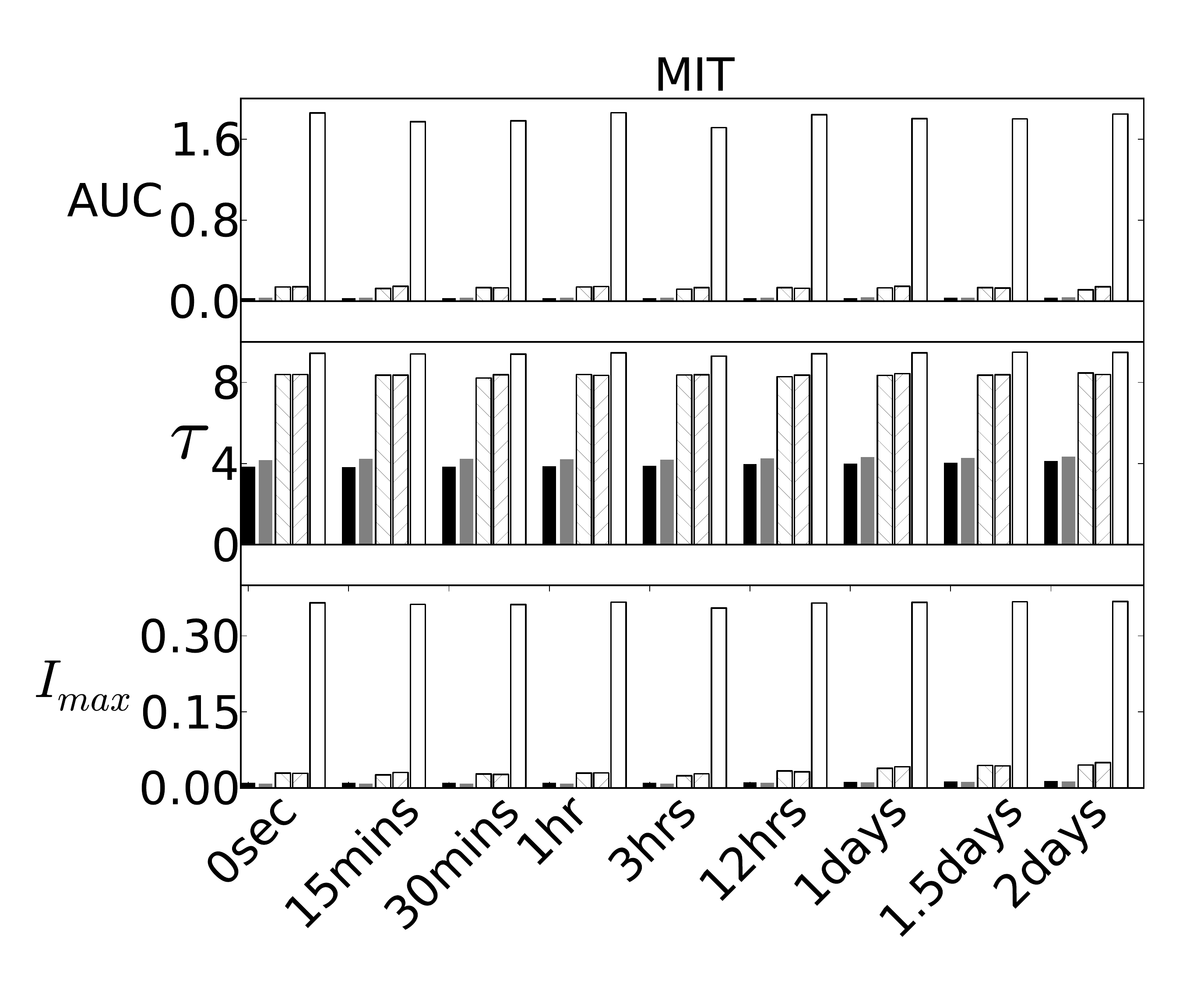}
	\end{minipage}
	\caption{Performance of temporal, static and naive node selection methods, as a function of patch delay (x-axis), averaged over all malware start times.}
	\label{fig:Iavg_N1_Pall}
\vspace{-12pt}
\end{figure*}
Now analysing the AUC and $I_{max}$, the temporal closeness centrality curve is consistently lower than static closeness, betweenness (both temporal and static) and naive methods. Further, betweenness (both static and temporal) generally take longer to fully contain the malware (higher values of $\tau$) and static closeness centrality performs worse than the naive method at some points of time; more specifically: 
\begin{itemize}
\vspace{-0.5pt} 
\item For the CAMBRIDGE dataset, during the weekend a static closeness method has a higher peak number of compromised devices ($I_{max}$) than the naive method, which shows that a static method is not effective at slowing down the malware from spreading.
\item For the INFOCOM dataset, again $I_{max}$ is higher than the naive method, during days 2 and 4. 
In addition, the AUC curve for a static method peaks with temporal efficiency during days 2, 4 and 5: this means that the malware is not contained effectively in these scenarios.  
Also, the total containment time ($\tau$) is greater than that of the naive method during days 3, 4 and 5. This shows that temporal closeness centrality is more consistent
at identifying the best nodes to start the patching process, compared to both static and naive methods.
\item Finally, for the MIT dataset, the naive method performs extremely poorly (with high values of AUC, $I_{max}$ and $\tau$ across all malware start times), compared to either a static or temporal methods.  
However, we also see that during the first week of the Fall semester, temporal closeness centrality identifies nodes with lower AUC and $\tau$, exhibiting over half a day quicker malware containment compared to static closeness centrality.
\end{itemize}
\subsubsection{Sensitivity to Patch Delay}
To understand the effects of delaying a patch message after a malware outbreak,  Figure~\ref{fig:Iavg_N1_Pall} plots the performance metrics for a representative sample of patch delays, averaged over all malware start times.  
As the patch delay increases, all the performance indicators also increase. 
However, we note that across all three datasets, temporal closeness centrality (left most bar) exhibits the best results: smallest AUC, fastest total containment time ($\tau$) and smallest peak compromised devices ($I_{max}$).  We also observe that in the INFOCOM dataset, static closeness node selection gives higher values of $I_{max}$ and $\tau$ up to a 12 hour delay, showing that static centrality does not consistently capture the true \textit{speed} at which a node can spread a message, compared to temporal closeness centrality.
Also, these plots demonstrate that betweenness (both static and temporal) are not suited to a spreading process and hence perform worse than closeness based node selection.
Again, from these observations, we conclude that a containment scheme based on temporal closeness centrality provides the best performance as the patch delay increases.

\subsection{Impact of the Initial Number of Compromised and Patching Devices}
\label{sec:increasing}
\begin{figure}[t!]
	\begin{minipage}[h!]{0.25\linewidth} 
		\centering
		\includegraphics[scale=0.2, trim = 3mm 0mm 0mm 0mm, clip]{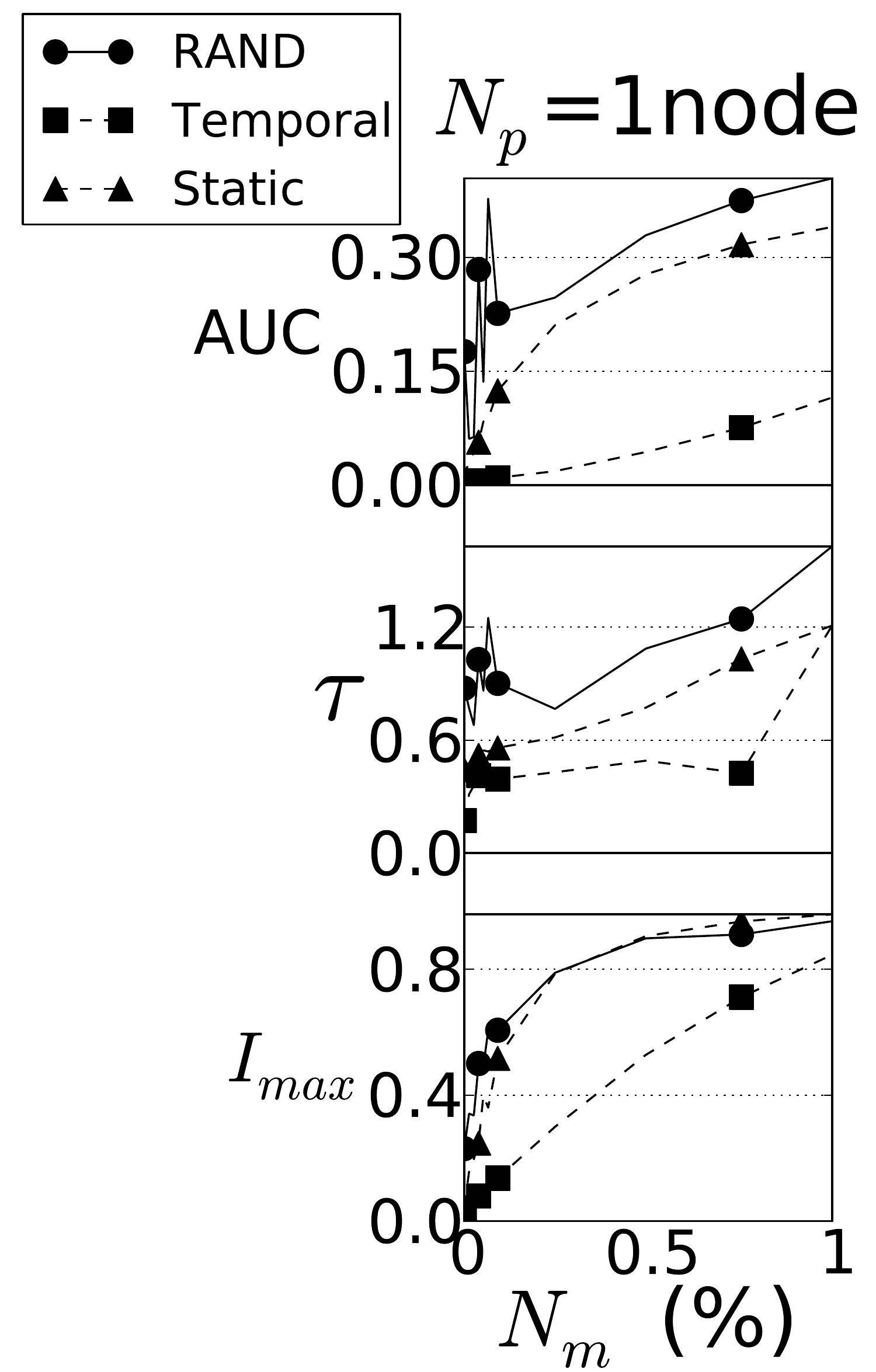}
	\end{minipage}
	\hspace{5.5mm} 
	\begin{minipage}[h!]{0.25\linewidth}
		\centering
		\includegraphics[scale=0.2, trim = 7.5cm 0mm 4mm 0mm, clip]{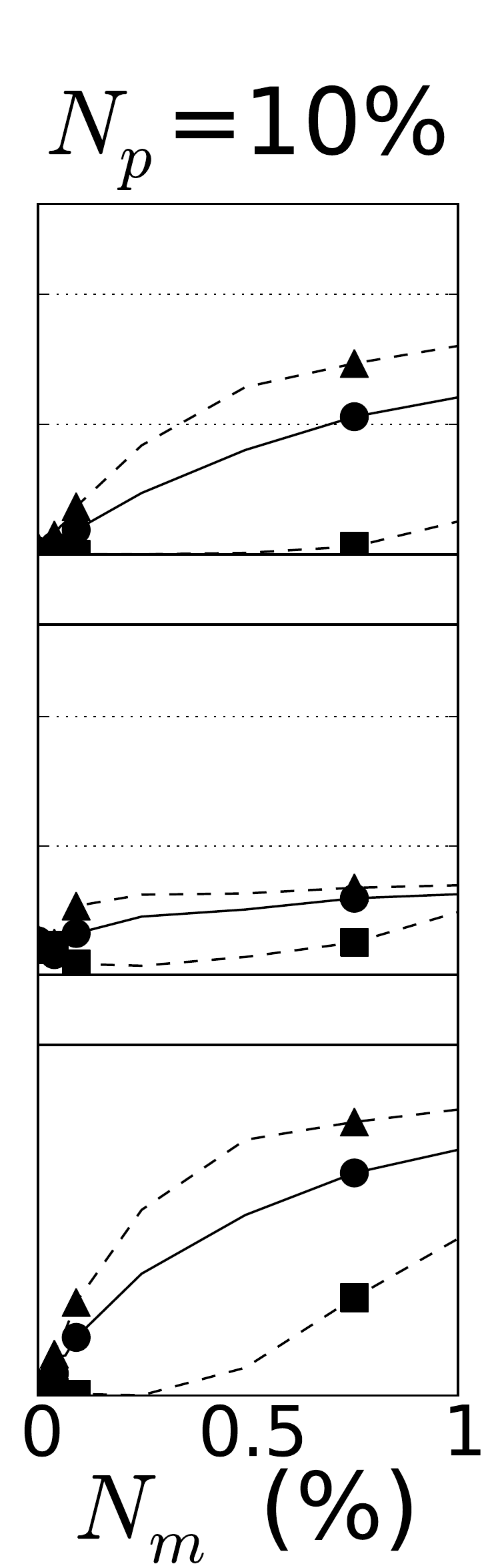}
	\end{minipage}
	\hspace{-6.0mm} 
	\begin{minipage}[h!]{0.25\linewidth}
		\centering
		\includegraphics[scale=0.2, trim = 7cm 0mm 4mm 0mm, clip]{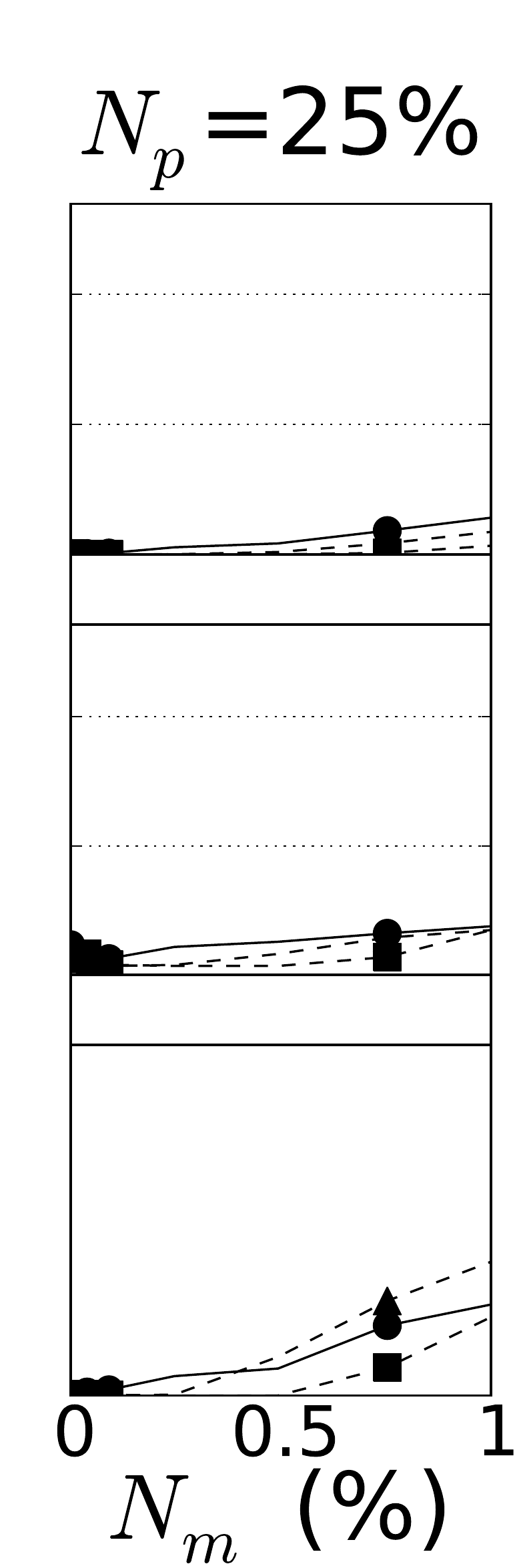}
	\end{minipage}
	\hspace{-6.0mm} 
	\begin{minipage}[h!]{0.25\linewidth}
		\centering
		\includegraphics[scale=0.2, trim = 7cm 0mm 4mm 0mm, clip]{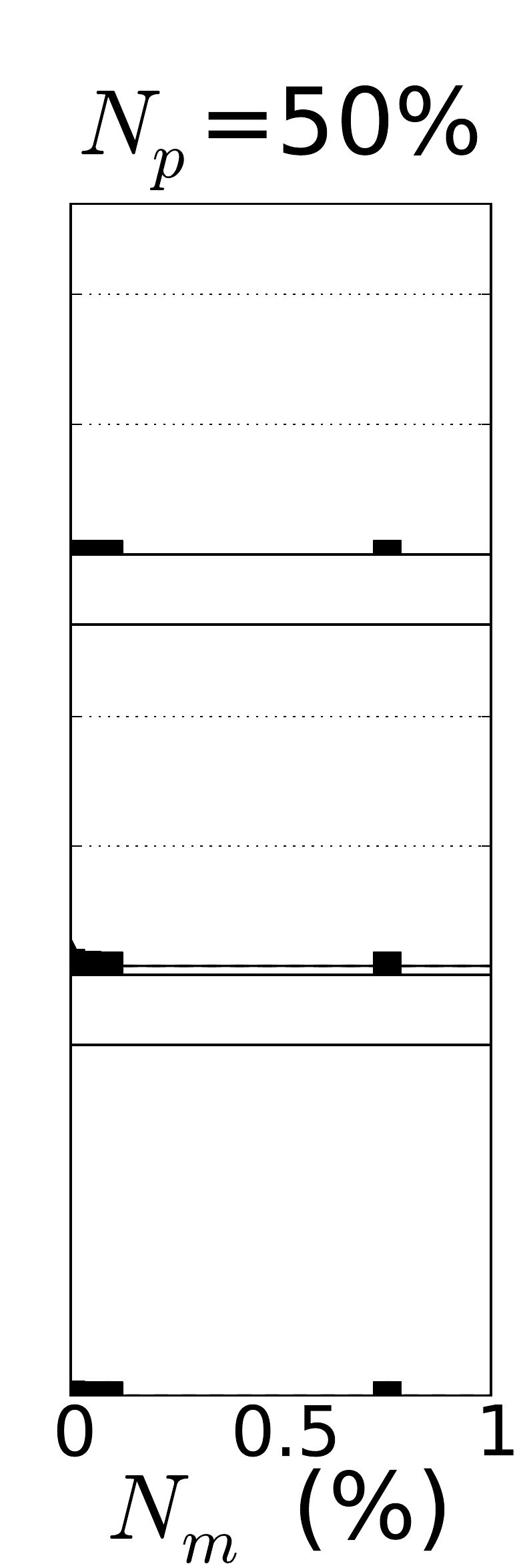}
	\end{minipage}
	\caption{INFOCOM: Effect of increasing number of initial devices with malware (x-axis).  From left to right, each column plots an increasing number of devices from which a patch is started ($t_m$=$t_p$=Day 4 12am).}\label{fig:infocon_Mall}
\vspace{-10pt}
\end{figure}

We now look at the effects of starting malware messages ($N_m$) and patch messages ($N_p$) from more than one device. 
This corresponds to the case, for example, when a group of people download a malicious program at the same time, or an attacker has programmed the replication to be time-triggered.  
Since we have observed that betweenness based node selection is not suited to patch spreading scheme, we now focus on closeness based node selection only. 
To make comparisons with the first containment scheme (Section \ref{sec:individual}) 
we discuss result for the same malware start and patch delay times.  Similar trends were found for different start times and other datasets.
Figure \ref{fig:infocon_Mall} shows the effect of starting a patch from an increasing number of initial devices $N_p$ 
(increasing column left to right) as the number of initially compromised devices $N_m$ (reported on the x-axis) is increased for the INFOCOM dataset.  

First, in the case when a single initial patch message ($N_p$=1) is used (left panel), we observe that the AUC corresponding to the scheme based on temporal centrality is lower than  that corresponding to the cases of static and naive methods of node selection even as $N_m$ increases; the total containment time ($\tau$) remains below half a day up to $N_m$=75\% of the total number of nodes (which we indicate with $N_{tot}$) and the peak compromised devices ($I_{max}$) rises slowly as $N_m$ increases.  
When increasing to $N_p$=10\%$N_{tot}$, using temporal centrality the total containment time ($I_{max}$) drops below 2.5 hours (about 0.1 of a day) up to $N_m$=75\%$N_{tot}$.  
Only at $N_p$=25\%$N_{tot}$ both the naive and static methods start to match the performance of the temporal method.  These observations suggest that our time-aware containment scheme using temporal centrality is more accurate at ranking important nodes and hence a viable option for a network operator since less devices are required to receive a patching message in order to achieve an effective containment strategy.  

\section{Discussion}
\subsection{Related Work}
\label{sec:relatedwork}

The study of techniques for containing the spreading of viruses and malware in the Internet has a long tradition (see a recent survey by Li \textit{et al.}~\cite{pele_li_survey_2008}). However, traditional desktop and server techniques for malware containment involve virus scanners running on a computer; such a scheme is not feasible on many mobile devices with limited resources.  More related to our work are so called ``white-worms" which propagate themselves in the same fashion as a malicious worm over the Internet, however, such Internet schemes have been documented as unsuccessful~\cite{weaver_white_2006}.  One reason for this is the speed at which a malicious worm can spread over the Internet relayed through computers and servers which are switched on 24 hours a day and connected to high speed networks; a white-worm has little chance of catching up.  Our results are positive because the basic assumption of connectivity in human contact networks is different: communication speed in mobile contact networks is not limited only by bandwidth but by actual physical proximity. As we have seen in Figure~\ref{fig:eff}, a malicious \textit{mobile} worm can only spread during times of high human activity, for example during the working day; this means that a mobile white-worm \textit{does} have time to catch up.
Our time-aware scheme adds on top of this a method for identifying the best devices to start the patch.

More recently, social network based strategies for the containment of mobile malware 
have been proposed.
In \cite{zhu_social_2009} Zhu \textit{et al.} propose that the most central nodes derived from phones call logs should be prioritised for patching.  However, this only captures potentially long-distance relationships and misses important opportunistic contacts that Bluetooth worms can exploit.  
In \cite{zyba_defending_2009} Zyba \textit{et al.} evaluate the spreading of a patch via short-range radio transmission; this work is based on a random mobility model and assumes homogeneous mixing and degree distribution over time.  As we have shown, mobile phone contact networks are driven by periodic human schedules and so the models proposed in this paper could be considered as an over-simplification of real situations. 

Such schemes are also partially founded on work on the robustness to random failures and targeted attacks of individual nodes in complex networks \cite{albert_error_2000}: these solutions are based on static graph representations which ignore time ordering and frequency of contacts. Instead, this work has shown that in real dynamic contact networks such schemes only \textit{slow} the spreading process and do not \textit{stop} it.

\subsection{Practical Implementation Issues}
In the same way that decentralised opportunistic routing protocols (such as BUBBLE Rap \cite{hui_bubble_2011} and SimBet \cite{daly_social_2009}) have utilised predictability in human contact networks through social network analysis to predict node centrality, 
our ongoing project will combine the time-aware malware containment strategy presented in this study with such heuristic and prediction techniques to identify the best devices to spread a patch at different times of the day.
Further, we shall investigate how such techniques can be enhanced by exploiting rich temporal information in human contact networks such as periodic behaviour \cite{clauset_2007} 
shown in Figure \ref{fig:eff}.

\section{Conclusions}
This paper has motivated and investigated the effectiveness of a time-aware mobile malware containment scheme using temporal centrality to identify the best node to start a competitive patch message.  The evaluation on three real human contact traces has shown that this time-aware scheme can more consistently identify the best devices to start such a patch across different malware start times and patch delays, compared to static and random 
node identification.

\section*{Acknowledgements}
The authors thank our shepherd Stratis Ioannidis 
 and the anonymous reviewers for their insightful comments which have helped in significantly improving our paper.  
 This work was supported by EPSRC Project MOLTEN (EP/I017321/1).

\begin{scriptsize}

\end{scriptsize}

\end{document}